\documentclass[aps,prx,twocolumn,english,superscriptaddress,floatfix,longbibliography]{revtex4-2}

\usepackage{amsfonts,amsmath,amssymb,amsthm}
\usepackage{array,bm}
\usepackage{epsfig,graphicx,nomencl,revsymb4-1,upgreek,url}
\usepackage{xspace}
\usepackage{float}
\usepackage{tabularx}
\newcolumntype{Y}{>{\centering\arraybackslash}X}
\usepackage{changes}
\usepackage{enumitem}
\usepackage{booktabs}
\usepackage{mathtools}

\usepackage{multirow}
\usepackage[colorlinks=true,hypertexnames=false,allcolors=blue]{hyperref}
\usepackage[utf8]{inputenc}

\usepackage[T1]{fontenc}
\usepackage{bm}
\usepackage[normalem]{ulem}

\usepackage{placeins}
\usepackage{qcircuit}
\usepackage[ruled,vlined,linesnumbered]{algorithm2e}
\usepackage{physics} 

\DeclareMathOperator*{\argmax}{arg\,max}

\newcommand{\fref}[1]{\hyperref[#1]{Fig.~\ref*{#1}}}
\newcommand{\Fref}[1]{\hyperref[#1]{Fig.~\ref*{#1}}}
\newcommand{\sref}[1]{\hyperref[#1]{Sec.~\ref*{#1}}}
\newcommand{\Sref}[1]{\hyperref[#1]{Sec.~\ref*{#1}}}
\newcommand{\tref}[1]{\hyperref[#1]{Table~\ref*{#1}}}
\newcommand{\Tref}[1]{\hyperref[#1]{Table~\ref*{#1}}}
\newcommand{\aref}[1]{\hyperref[#1]{Appendix~\ref*{#1}}}
\newcommand{\Aref}[1]{\hyperref[#1]{Appendix~\ref*{#1}}}
\newcommand{\thref}[1]{\hyperref[#1]{Theorem~\ref*{#1}}}
\newcommand{\Thref}[1]{\hyperref[#1]{Theorem~\ref*{#1}}}

\newcommand{\loschmidtf}{F_\text{LE}}
\newcommand{\loschmidtfon}[1]{\loschmidtf\left(#1\right)}
\newcommand{\noisyop}[1]{\mathcal{E}_{#1}} 
\newcommand{\noisyopon}[2]{\noisyop{#1}\left(#2\right)} 
\newcommand{\zerodm}{|0\rangle\langle 0|}
\newcommand{\I}{\mathbb{I}}

\usepackage[normalem]{ulem}

\newcommand{\delete}[1]{}
\newcommand{\edit}[1]{#1}
\definecolor{vermillion}{rgb}{0.86, 0.18, 0.01}

\newcommand{\evan}[1]{}
\newcommand{\prasanth}[1]{}

\theoremstyle{definition}

\newtheorem{theorem}{Theorem}
\newtheorem*{theorem*}{Theorem}

\newtheorem*{lemma*}{Lemma}

\newtheorem*{corollary*}{Corollary}
\newtheorem{proposition}[theorem]{Proposition}

\newcommand{\reportnumber}{FERMILAB-PUB-21-726-QIS}
\usepackage[angle=0,scale=1,color=black,firstpage=true,opacity=1]{background}

\SetBgContents{\small\texttt{\reportnumber}}
\SetBgPosition{current page.north east}
\SetBgHshift{-3.5cm}
\SetBgVshift{-1.5cm}

\begin{document}

\title{
Qubit assignment using time reversal
}
\author{Evan Peters}
\email{e6peters@uwaterloo.ca}
\affiliation{Fermi National Accelerator Laboratory, Batavia, IL 60510}
\affiliation{Institute for Quantum Computing, University of Waterloo, Waterloo, Ontario, N2L 3G1, Canada}
\affiliation{Department of Physics, University of Waterloo, Waterloo, Ontario, N2L 3G1, Canada}
\author{Prasanth Shyamsundar}
\affiliation{Fermi National Accelerator Laboratory, Batavia, IL 60510}
\author{Andy C.~Y.~Li}
\affiliation{Fermi National Accelerator Laboratory, Batavia, IL 60510}
\author{Gabriel Perdue}
\email{perdue@fnal.gov}
\affiliation{Fermi National Accelerator Laboratory, Batavia, IL 60510}

\date{October 8, 2022}

\begin{abstract}

As quantum computers with large numbers of qubits become increasingly available, experiments executed on a given device may not utilize all available qubits. In this case, the outcome of executing a quantum program will depend on the ability to efficiently select a subset of high-performing physical qubits. For any given quantum program and device there are many ways to assign physical qubits for execution of the program, and assignments will differ in performance due to the variability in quality across qubits and entangling operations on a single device. Evaluating
the performance of each assignment using fidelity estimation introduces significant experimental overhead and
will be infeasible for many applications, while relying on standard device benchmarks provides incomplete information about the performance of any specific program. Furthermore, the number of possible assignments grows
combinatorially in the number of qubits on the device and in the program, motivating the use of heuristic optimization techniques.
We demonstrate a practical solution to the problem of qubit assignment by using simulated annealing with a cost function based on the Loschmidt
Echo, a diagnostic that measures the reversibility of a quantum process. We provide theoretical justification for this choice of cost function by demonstrating that
the optimal qubit assignment coincides with the optimal qubit assignment based on state fidelity in the weak error limit, and we provide experimental justification using diagnostics performed on Google’s superconducting qubit devices. We then establish the performance of simulated annealing for qubit assignment using classical simulations of noisy devices as well as optimization experiments performed on a quantum processor. Our results demonstrate that the use of Loschmidt Echoes and simulated annealing provides a scalable and flexible approach to optimizing qubit assignment on near-term hardware.

\end{abstract}




\maketitle

\section{Introduction}

With the increased availability of NISQ \cite{Preskill_2018} devices there is a growing need for tools to efficiently deploy quantum circuits on hardware in a noise-aware manner. Part of this task is qubit assignment \cite{finigan_qubit_2018,tannu_not_2019}, where the goal is to assign a \edit{logical circuit \footnote{In the context of this work, a ``logical circuit'' is an abstraction used in the specification of a sequence of gates acting on qubits that have no correspondence to any physical device. Meanwhile, ``physical qubits'' exist on a specific device and are subject to connectivity constraints and errors. Qubit assignment is the process of selecting a correspondence between each logical qubit and some physical qubit that respects connectivity constraints. This differs from the logical-vs-physical distinction typically used in the context of quantum error correction (QEC).}}  to the set of physical qubits on noisy hardware that maximizes the circuit performance. This task requires access to a performance metric for circuits implemented on hardware that is both efficient to evaluate and faithful to standard fidelity metrics, as well as a means of efficiently optimizing that performance metric with respect to the set of all possible qubit assignments.

Performing qubit assignment in a way that satisfies these requirements faces a number of challenges. For instance, an experimentalist might choose physical qubits according to which subset maximizes the fidelity of the output of the quantum program, but this introduces an experimental overhead that is exponential in the system size on near term devices. On the other hand, choosing physical qubits based on standard device benchmarks such as randomized benchmarking \cite{Emerson_2005,Knill_2008} or cross-entropy benchmarking \cite{Boixo_2018,Neill_2018} can result in poor-performing assignments since these benchmarks capture average error behavior that may differ from the noise occurring in the context of a specific quantum program. Once a performance metric is chosen, there still remains the challenge of efficiently exploring a space of qubit assignments that grows combinatorially both in the size of the quantum program and the hardware device.
\edit{Implementing qubit assignment techniques can improve device performance whenever the total number of qubits on the device is larger than the number required for an application.
This scenario is typical for near and intermediate-term devices, particularly in the context of quantum error correction experiments \cite{Chen2021,McEwen2021,PhysRevLett.128.110504,PhysRevLett.129.030501,https://doi.org/10.48550/arxiv.2207.06431}.}

To overcome these challenges we study the Loschmidt Echo \cite{PhysRevA.30.1610}, a tool for probing reversibility in quantum systems. We then demonstrate that this metric can be used with simulated annealing (SA) \cite{van1987simulated} to effectively perform qubit assignment on hardware. The large state space of potential hardware circuits combined with high variance in qubit error rates naturally leads to an uneven cost landscape, for which SA is particularly well suited. Combining SA with Loschmidt Echoes then provides a scalable technique for qubit assignment that does not rely on potentially inaccurate hardware diagnostics.

Prior works have used variants of the Loschmidt Echo to assess the performance of qubit assignments. Refs. \cite{proctor2020measuring,proctor2021scalable}  introduced techniques for benchmarking quantum devices using ``mirror circuits'' composed entirely of Clifford operations, for which the fidelity of output state may be computed with the aid of efficient classical simulations of Clifford circuits \cite{gottesman1998heisenberg,gottesman_aaronson_2004}. While this technique provides a benchmark for the performance of a set of qubits with respect to a general set of operations, we are interested in the ability of the device to accurately prepare some specific state $|\psi\rangle$. \edit{Furthermore, while it is possible to use mirror circuits to evaluate the performance associated with specific circuits, its application to large circuits is limited due to the lack of efficient classical simulation of general non-Clifford circuits.} Moreover, we will demonstrate \edit{theoretically how generic benchmarks (which assess qubit and gate performance independently of circuit structure) have limited ability to predict the performance of a given qubit assignment. We will provide experimental evidence to support this claim for a specific choice of generic benchmark.}

Similarly, our work differs from prior qubit assignment experiments \cite{finigan_qubit_2018,Nishio_2020} since we do not rely on a set of gate fidelities characterizing hardware performance as a proxy for assessing the performance of a specific circuit on hardware. Therefore, while Ref. \cite{finigan_qubit_2018} used a hybrid algorithm involving SA to search over ``sub-allocations'' of progressively larger qubit subsets, in our approach we altogether avoid using a graph weighted by gate fidelities of (partial) circuits to anneal over.

Qubit assignment \cite{maslov2008,childs2019circuit,10.1145/3316781.3317859,tan2020optimal,gerard2021string} (often referred to as qubit routing, qubit allocation, or quantum compilation) has been extensively studied as a tool for improving the performance of circuits executed on hardware. Qubit assignment typically includes modifying logical circuits to run on hardware when the gateset and connectivity constraints of the device do not match those of the logical program, typically with the goal of minimizing the number of additional operations introduced to the program. For instance, Ref. \cite{Zhou_2020} introduced a technique for circuit compilation using SA with a cost function based on CNOT count. However, in this work we simplify qubit assignment on hardware by only considering a fixed circuit that already satisfies hardware connectivity, with the goal of maximizing the performance of a quantum program with respect to the choice of hardware qubit subsets satisfying this connectivity. 

This work proceeds as follows. In Sec.~\ref{sec:graphstuff} we overview the problem of qubit assignment on a hardware device. In Sec.~\ref{sec:loschmidt_echo} we provide theoretical evidence that the Loschmidt Echo is useful for ranking qubit assignments when the goal is to maximize the fidelity of a prepared state, and demonstrate shortcomings of other existing methods based on benchmark data or random circuit fidelities in this same context. Specifically, we show that scoring qubit assignments based on gate fidelities taken from device benchmark data fails to capture general performance trends. In Sec.~\ref{sec:SA_circuit} we then develop a framework for performing qubit assignment using SA, and in Sec~\ref{sec:experiments} we verify the performance of the Loschmidt echo and demonstrate optimization results that outperform competitive techniques on both simulated and experimental datasets.

\section{Methods}\label{sec:methods}
\subsection{Qubit assignment on hardware}\label{sec:graphstuff}

We describe the task of qubit assignment in terms of a hardware topology and a set of gates \cite{10.1145/3168822}. We will provide a graph-based description of a quantum circuit as a sequence of gates applied to single qubits or pairs of qubits in a logical circuit, and a representation of qubit assignment as a transformation subject to connectivity constraints on a hardware device.

We describe the connectivity of a quantum device by an undirected graph $G_p = (V_p, E_p)$ where each vertex $i \in V_p$ describes a physical qubit and each edge $\{i, j\}\in E$ indicates that entangling operations can be executed between qubits $i$ and $j$. We assume $|V_p|=N$ qubits are available on hardware and that the hardware supports a gateset $\mathcal{G}_p$ consisting of single- and two-qubit gates which allows for universal computation \cite{kitaev1997} and is a common choice on superconducting qubit platforms \cite{qiskit,cirq}. We similarly define a logical graph $G_L=(V_L, E_L)$ and a logical circuit $\mathcal{C}_L$ over $n$ qubits as a sequence of $m$ gate operations $[g_1(v_1, e_1), \dots, g_m(v_m, e_m)]$, with $e_k\in E_L$, $v_k \in V_L$, and $g_k\in \mathcal{G}_L$ for $k=1 \dots m$. Taking every gate $g_k$ in $\mathcal{C}_L$ to act on either $(i, \emptyset)$ (single-qubit gate) or $(\emptyset, \{i, j\})$ (two-qubit gate), we can summarize the logical circuit $\mathcal{C}_L\in  (\mathcal{G}_L^m, V_L^m, E_L^m)$ according to three sequences, one for each gate description, one for each set of target qubits, and one for each set of target edges. For example a circuit might look like
\begin{align}
\begin{matrix}
     \textbf{g} &= &\\
     \textbf{v}_L &= &\\
     \textbf{e}_L &= &
\end{matrix} 
\begin{matrix}
     [& R_x(\theta),& \text{CNOT},& X,& \dots]\\ 
     [& \{3\},& \emptyset,& \{3\},& \dots]\\
     [& \emptyset,& \{1, 4\},& \emptyset,& \dots]
\end{matrix}
\end{align}


Let $U(\mathcal{C}_L)$ be the unitary representation of $\mathcal{C}_L$ acting on $n$ unique qubits and and let $U(\mathcal{C}_p)$ be the unitary representation of a hardware program $\mathcal{C}_p \in (\mathcal{G}_p^{m'}, V_p^{m'}, E_p^{m'})$ acting on up to $N\geq n$ qubits. Then defining the state $|\psi\rangle= U(\mathcal{C}_L) |0^n\rangle$ we are interested in searching over the range of a logical to physical circuit map of the form
\begin{equation}\label{eq:simple_map}
M: (\mathcal{G}_L^m, V_L^m, E_L^m) \rightarrow (\mathcal{G}_p^{m'}, V_p^{m'}, E_p^{m'})
\end{equation}
subject to the constraint
$$|\psi\rangle\langle\psi| = \Tr_{A}\left(U(\mathcal{C}_p)|0^N\rangle \langle 0^N | U(\mathcal{C}_p)^\dagger \right)$$

\noindent for some subsystem $A$ consisting of up to $N - n$ qubits, in which case we will say that $M$ satisfies $M(\mathcal{C}_L) = \mathcal{C}_p$. In general, the effects of noise on the quantum device will interfere with perfect realization of the unitary operation $U(\mathcal{C}_p)$. We therefore assume that there is a quantum channel $\mathcal{E}(\rho; \mathcal{C}_p)$ that describes the effect of $\mathcal{C}_p$ applied to an input state $\rho$ on hardware. We will consider implementing a specific unitary $U$ on noisy hardware and therefore restrict our attention to circuits $\mathcal{C}_p$ for which $U(\mathcal{C}_p) = U$ up to some permutation of \edit{subsystems and} we will hereafter implicitly assume dependence on some hardware circuit $\mathcal{C}_p$ (and therefore some qubit assignment) using the notation $\noisyopon{U}{\rho} = \mathcal{E}(\rho; \mathcal{C}_p)$.

\subsection{Loschmidt Echoes for evaluating qubit assignments}\label{sec:loschmidt_echo}

We are interested in selecting the best qubits for preparing a state $|\psi\rangle = U|0\rangle$ over $n$ qubits given a choice of $N$ hardware qubits with constrained topology.
Ideally we would be able to directly compare the state prepared by a noisy implementation $\mathcal{E}_U$ to the perfect preparation of $|\psi\rangle = U\ket{0}$. One way to characterize the hardware performance is to employ the  fidelity function \cite{nielsen2002quantum}, which provides a measure of closeness between two quantum states $\rho$ and $\sigma$ given as $\mathcal{F}(\rho, \sigma)=\left(\Tr \sqrt{\rho^{1/2} \sigma \rho^{1/2}}\right)^2$. Using $\mathcal{F}$ we can compare the prepared state $\rho = \noisyopon{U}{\zerodm}$ (which implicitly depends on a hardware circuit $\mathcal{C}_p$) to an ideal state prepared by $U$ in a noiseless environment:
\begin{align}\label{eq:true_fid}
    F(\mathcal{C}_p) :=  \mathcal{F}(\rho, |\psi\rangle\langle \psi|)
    =\langle \psi| \rho |\psi\rangle 
\end{align}

\begin{figure}
    \centering
    \begin{minipage}{0.45\textwidth}
        \[ 
        \Qcircuit @R=1em @C=0.75em {
         \\
         &\lstick{|0^n\rangle}& \qw &\gate{U} \qw &\gate{U^\dagger} \qw& \qw&\measureD{0^n} \qw&\cw\\
        }
        \]
    \end{minipage}\hfill
    \begin{minipage}{0.45\textwidth}
        \[ 
        \Qcircuit @R=1em @C=0.75em {
         \\
         &\lstick{|0^n\rangle\langle 0^n|}& \qw &\measure{\noisyop{U}} \qw &\measure{\noisyop{U^\dagger}} \qw& \qw&\measureD{0^n} \qw&\cw\\
        }
        \]
    \end{minipage}
    \caption{(top) Logical circuit for sampling $|\langle 0| U^\dagger U |0\rangle|^2$, which will equal unity in the absence of device error. (bottom) schematic representation for sampling $\loschmidtfon{\mathcal{C}_p} = \langle 0| \noisyop{U^\dagger} \circ \noisyopon{U}{\zerodm} |0\rangle$, where curved edges indicate a completely positive trace preserving (CPTP) map acting on $n$-qubit density operators.}
    \label{fig:loschmidt_circuit}
\end{figure}
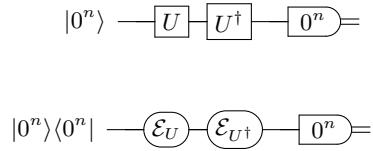

In general, evaluating $F$ on a near term device can incur significant resource overhead. Computing $F$ using Direct Fidelity Estimation (DFE) \cite{PhysRevLett.107.210404,PhysRevLett.106.230501} can  incur significant experimental overhead for each choice of qubits, and is therefore impractical to evaluate for qubit assignment on large quantum devices. Similarly, classical shadows \cite{huang2020predicting} can be used evaluate $F$ in constant time, but evaluating \edit{fidelities in this way} requires significant overhead and the ability to execute operations drawn randomly from the $n$-qubit Clifford group --- a requirement that is out of reach for near-term devices due to the depth requirements of such circuits. We are therefore motivated to find a proxy for circuit fidelity that can be executed quickly and efficiently, at the expense of accurately characterizing the true state fidelity or process fidelity. Given the ability to implement unitaries $U$ and $V$, the fidelity between the states $|\psi\rangle = U|0\rangle$ and $|\phi\rangle = V|0\rangle$ can be determined in constant time since the quantity
\begin{align}\label{eq:state_fid}
    \mathcal{F}(|\psi\rangle\langle \psi|, |\phi\rangle\langle \phi|) = |\langle \psi |\phi \rangle |^2 
    = |\langle 0 | U^\dagger V |0\rangle|^2
\end{align}
can be estimated as the empirical probability of observing the all-zeros bitstring $0^n$ at the output of a circuit $U^\dagger V$.
By extension, to test the effects of noise we choose to evaluate a performance metric for $\mathcal{C}_p$ defined as

\begin{equation}\label{eq:roundtrip_fid}
    \loschmidtfon{\mathcal{C}_p} = \Tr \left( \zerodm  \noisyop{U^\dagger} \circ \noisyop{U}\zerodm \right)
\end{equation}

where $\mathcal{E}_{U^\dagger}$ represents the process resulting from a noisy implementation of a unitary $U^\dagger$ on hardware. Eq.~\ref{eq:roundtrip_fid} computes the probability that a state prepared by implementing $U$ can be successfully returned to $|0\rangle$ by implementing $U^\dagger$, though this will not provide an accurate estimate for the state fidelity $F$ in general. This process is  referred to as a Loschmidt Echo  \cite{loschmidt1876,PhysRevA.30.1610}, and serves as a general measure for reversibility in a quantum process (see Ref. \cite{loschmidt_review} for a thorough review). In the special case where $\noisyop{U^\dagger}(\rho) = U^\dagger \rho U$ can be implemented with perfect fidelity, comparison to Eq.~\ref{eq:true_fid} shows that this procedure exactly recovers $F(\mathcal{C}_p)$. From this perspective, the application of $\noisyop{U^\dagger}$ can be understood as preparing an imperfect measurement of the projector $|\psi\rangle\langle \psi|$, a comparison that suggests that $\loschmidtf$ may be an effective tool for analyzing the implementation of $U$ on hardware. Fig.~\ref{fig:loschmidt_circuit} shows a circuit which estimates $\loschmidtf$ as the probability of sampling the all-zeros bitstring at the circuit output.

While $\loschmidtf$ will generally not provide an estimate for the state fidelity $F$, we now show that given some assumptions about device noise, the optimal qubit assignment with respect to maximizing $\loschmidtf$ will result in a state with high fidelity as computed by Eq.~\ref{eq:true_fid}. We first provide a concrete statement of this result by assuming a noise model consisting of one-qubit and two-qubit local depolarizing channels parameterized by the error rate $\epsilon_i$ for each single-qubit gate acting on qubit $i$ and the error rate $\eta_{ij}$ for each two-qubit gate acting on qubits $\{i, j\}$.  This model is applied to the hardware implementation of each hardware circuit $\mathcal{C}_p' = (\textbf{g}, \textbf{v}', \textbf{e}')$ with a fixed gate sequence $\textbf{g}$ of length $m$. We then define the sets of optimal qubit assignments with respect to $F$ and $\loschmidtf$ respectively as:
\begin{align}
    \label{eq:Qstar}
    S^* &= \{ S': S' = \argmax_{\mathcal{C}_p':\, U(\mathcal{C}_p') = U} F(\mathcal{C}_p')\} \\
    \label{eq:QLE}
    S_{LE}^* &= \{ S': S' = \argmax_{\mathcal{C}_p':\, U(\mathcal{C}_p') = U} \loschmidtfon{\mathcal{C}_p'}\}
\end{align}
where the fixed gate sequence means that each maximum is computed with respect to all possible, valid assignments on the hardware device, i.e. sequences of qubits $\textbf{v} \in (V')^{m}$ and sequences of edges $\textbf{e} \in (E')^{m}$ subject to  $V' \subset V_p$ and $E' \subset E_p$ for a device with connectivity described by $(V_p, E_p)$. Under the assumed noise model, we can then compare these optimal assignments in a low-error limit by the following proposition:

\begin{proposition}\label{prop:1}
For a graph $G_p = (V_p, E_p)$ describing hardware qubits, consider an error model consisting of \edit{Markovian noise channels, each characterized by an error rate $\epsilon$, inserted after each gate. Suppose each gate $U$ and its inverse implementation $U^{\dagger}$ are augmented by the same noise channel}.
\delete{$n_i$ many single-qubit depolarizing channels with error probability $\epsilon_i$ inserted after each single-qubit gate acting on qubit $i\in V_p$, and $n_{ij}$ many two-qubit symmetric depolarizing channels with error probability $\eta_{ij}$ inserted after each two-qubit gate acting on qubits $\{i, j\} \in E_p$.}
Let the optimal qubit assignments $S_{LE}^*$ and $S^*$ be defined as in Eqs.~\ref{eq:Qstar}-\ref{eq:QLE}. 
Then in the weak-error limit \edit{$\epsilon \ll 1$} \delete{$(\epsilon_i n_i) << 1$, $(\eta_{ij} n_{ij}) << 1$ for all $i,j \in V_p$ and $\{i,j\}\in E_p$},
\begin{equation}
    S_{LE}^* = S^*
\end{equation}
\end{proposition}

\edit{We first illustrate this claim by considering the special case of local depolarizing noise (Appendix~\ref{app:depolarizing_noise_channel}) and global depolarizing noise (Appendix~\ref{app:other_noise}), followed by a complete proof of this proposition  then} in Appendix~\ref{app:markovian}.
A simple counterexample involving unitary errors for which $\loschmidtf$ is independent of $F$ is also provided in Appendix~\ref{app:other_noise}.
\delete{However we also emphasize the importance of our choice of noise model by demonstrating simple counterexamples involving unitary errors for which $\loschmidtf$ is independent of $F$ (Appendix~\ref{app:other_noise}). While Proposition~\ref{prop:1} is presented in terms of a specific noise model that is convenient for comparison to standard device benchmarks, it is actually a special case of a more general analysis that can be applied to $\loschmidtf$ under certain conditions imposed on the noise model.
In Appendix~\ref{app:markovian} we demonstrate that $S_{LE}^* = S^*$ also holds in the weak-error limit in any system where the environmental noise is Markovian and the coherent error remains the same for any gate $U$ and its inverse implementation $U^{\dagger}$.}
In Appendix~\ref{app:readout_error} we show that $\loschmidtf$ is strongly biased by readout error compared to the true fidelity of the prepared state, and so it is necessary to implement readout error mitigation \cite{Neeley_2010,Dewes_2012,nachman_unfolding_2020,bravyi2020mitigating,PRXQuantum.2.040326} and in particular, readout correction to recover the all-zeros bitstring \cite{peters2021perturbative} when selecting qubits based on this metric.

Under the assumption of \edit{a 2-local depolarizing error model discussed in Appendix~\ref{app:depolarizing_noise_channel}} \delete{the above error model}, it is natural to consider simpler metrics that might be used to estimate the performance a qubit assignment without the need for circuit-specific hardware experiments. For instance, the fidelity of random circuits exposed to discrete errors can be well approximated using device diagnostic data \cite{arute2019quantum}, and so in addition to computing $\loschmidtf$ and $F$ for each qubit assignment, we also consider a heuristic that can be computed by combining the structure of the executed circuit with device benchmark data:
\begin{equation}\label{eq:F0}
    F_0(\boldsymbol{\epsilon}, \boldsymbol{\eta}) = \prod_{i\in V_p} ( 1-\epsilon_i)^{n_i} \prod_{\{i,j\}\in E_p} ( 1-\eta_{ij})^{n_{ij}} 
\end{equation}
where $n_i$ counts the number of single qubit gates acting on qubit $i$, $n_{ij}$ counts the two-qubit gates acting on edge $\{i,j\}$, the length-$|V_p|$ vector $\boldsymbol{\epsilon}$ contains weights for the graph vertices, and the length-$|E_p|$ vector contains weights for the graph edges. These weights describe error rates for single-qubit gates and two-qubit gates respectively. We omit a term characterizing measurement error from $F_0$ since we assume that readout error correction will be employed when estimating $F$ and $\loschmidtf$. 

Assuming the 2-local depolarizing noise model \delete{of Proposition~\ref{prop:1}}, $F_0$ is the probability that an error does not occur during the execution of $U$ on noisy hardware with intrinsic polarization error rates defined by $\boldsymbol{\epsilon}$ and $\boldsymbol{\eta}$.
This suggests that an experimentalist could perform device diagnostics such as randomized benchmarking (RB) \cite{Emerson_2005,Knill_2008} or cross-entropy benchmarking (XEB) \cite{Boixo_2018,Neill_2018} to estimate the gate error rates as $\hat{\boldsymbol{\epsilon}}$ and $\hat{\boldsymbol{\eta}}$ and then compute $F_0(\hat{\boldsymbol{\epsilon}}, \hat{\boldsymbol{\eta}})$ to evaluate the performance of any given qubit assignment using Eq.~\ref{eq:F0}.

This strategy has been used in previous works for qubit assignment \cite{finigan_qubit_2018,murali2019noise,peters2021machine} (the approaches of Refs. \cite{niu2020,8980312} involved finding a shortest path on a graph with weights $\{\eta_{ij}\}$, though the operational meaning of a sum of error rates is not as clear compared to the product formula for $F_0$). \edit{However we argue that} this strategy must be approached with caution since it neglect information about how device noise affects a specific circuit. Even in the scenario of a circuit exposed to depolarizing noise where an experimentalist is given perfect knowledge of depolarizing error rates, we show in Appendix~\ref{app:prop1} that the optimal qubit assignment with respect to $F_0$ will not necessarily maximize $F$. Errors in experimental characterization of device performance (e.g. due to drift \cite{PhysRevX.9.021045}) will then only compound this limitation. Our experimental results in Sec.~\ref{sec:hardware} will demonstrate that $F_0$ is almost useless as a proxy for determining the \edit{relative performance of qubit assignments on actual hardware}. 

In our experiments we also consider performing qubit assignment with respect to an average Loschmidt survival rate $\langle F_{LE}^{rand}\rangle$ computed over a series of random circuits. Random circuits have been used to successfully evaluate the average fidelity of operations in theory and on quantum hardware  \cite{arute2019quantum,Knill_2008,Magesan_2011,Magesan_2012,liu2021benchmarking}, and therefore $\langle F_{LE}^{rand}\rangle$ will be sensitive to coherent errors that may not be captured by $\loschmidtf$. Our results will show that optimizing a qubit assignment with respect to $\langle F_{LE}^{rand}\rangle$ will serve as a weak proxy for determining the optimal qubit assignment for a circuit $U$, unless $U$ is itself a random circuit.
\subsection{Simulated annealing for qubit selection}\label{sec:SA_circuit}

Simulated Annealing (SA) \cite{van1987simulated} is an optimization technique motivated by the behavior of physical systems to approach their ground state when cooled sufficiently slowly. We now describe the task of SA on finite sets \cite{bertsimas1993simulated} in the context of qubit assignment.
Given a fixed logical circuit $\mathcal{C}_L$ to execute and the set $S = \{\mathcal{C}_p: M(\mathcal{C}_L)=\mathcal{C}_p\}$ of possible hardware circuits to evaluate, accompanied by a cost function $C: S\rightarrow \mathbb{R}$ to be minimized, our goal is to find the optimal subset $S^* \subseteq S$ corresponding to the minimum value of $C$.
For hardware experiments we make a number of assumptions about $M$ that guarantee that $S$ will be finite. Namely, we restrict the sequence of gates $\mathcal{C}_p$ to be identical to the sequence of gates in $\mathcal{C}_L$ and fix the logical connectivity to be the same as the physical device connectivity, so that $M$ can be understood as a one-to-one map between logical and physical qubits (see Appendix~\ref{app:qubit_set_discussion}). For a physical circuit $\mathcal{C}_p$ implementing a corresponding noisy process $\noisyop{U}$, we use the Loschmidt Echo performance metric of Sec.~\ref{sec:loschmidt_echo} to define a cost function as 
\begin{equation}\label{eq:cost_def}
    C(\mathcal{C}_p) = 1 - \loschmidtfon{\mathcal{C}_p}
\end{equation}
where $\loschmidtf$ is implicitly related to the set of qubits and entangling edges output by $M$. By analogy with annealing in physical systems, we define a temperature $T(t)$ at each iteration $t$ of the algorithm that dictates how likely the annealer is to evaluate a worse state in the interest of exploration. Then, given a choice between a current physical circuit $\mathcal{C}_p$ and a new circuit $\mathcal{C}_p'$ at each step, the algorithm will probabilistically select $\mathcal{C}_p'$ depending on $C(\mathcal{C}_p')$ and $T$: If $C(\mathcal{C}_p') < C(\mathcal{C}_p)$ then the algorithm always chooses $\mathcal{C}_p'$. Otherwise if $C(\mathcal{C}_p') > C(\mathcal{C}_p)$, SA will still allow for exploration of the state space with probability
\begin{equation}
    P(\mathcal{C}_p \rightarrow \mathcal{C}_p') = \exp\left( \frac{C(\mathcal{C}_p ') - C(\mathcal{C}_p ) }{T(t)}\right)
\end{equation}

and will therefore be more likely to explore disadvantageous configurations in the early stages of exploration to avoid becoming trapped in a local minumum. The implementation of SA for qubit assignment depends strongly on the set of new circuits that the algorithm can sample from at each step. Given a physical circuit $\mathcal{C}_p$, at any given iteration we define the neighborhood of circuits $S(\mathcal{C}_p)$ at $S$ as the set of all circuits $\mathcal{C}_p'\neq \mathcal{C}_p$ that the algorithm may sample from. The definition of a neighborhood is important for understanding and tuning the convergence properties of finite SA. If the set of neighbors is too large, such as the extreme example $S_{rand}(\mathcal{C}_p) = S \,\backslash \,\{\mathcal{C}_p\}$ \edit{(i.e., $S_{rand}$ contains all circuits in $S$ except $\mathcal{C}_p$)} then SA performs a random walk over $S$, and convergence to $S^*$ requires the maximum number of iterations over all possible choices of $S(\mathcal{C}_p)$ \cite{hajek1988cooling}. Conversely if the neighborhood is too small, then SA may require a prohibitive runtime to reach $S^*$. For each experiment we will implicitly or explicitly define $S(\mathcal{C}_p)$ as it relates to the logical circuit connectivity. In~Appendix~\ref{app:annealer_algorithms} we provide Algorithm~\ref{alg:sa_finite} with pseudocode to implement SA for qubit assignment.


\subsection{Logical and physical connectivity}\label{sec:constrained_1}

For hardware experiments on a device containing a grid of qubits with nearest neighbor connectivity, we consider logical circuits with nearest neighbor line connectivity, thereby restricting $S$ and $S(\mathcal{C}_p)$. Each circuit $\mathcal{C}_L = (\mathbf{g}, \mathbf{v}_L, \mathbf{e}_L)$ to is defined over $n$ logical qubits with edges between nearest neighbors:
\begin{align}\label{eq:line1}
     \mathbf{v}_L&= [v_1, \dots, v_n] \\
     \mathbf{e}_L&=[\{v_1, v_2\}, \dots, \{v_{n-1}, v_n\} ]
\end{align}
where $v_k \in V_p\text{ for }k=1\dots n$ and $\{v_\ell, v_{\ell+1}\} \in E_p\text{ for } \ell=1\dots n-1$  subject to the restriction that $\{v_1, \dots, v_n\}$ compose a length-$n$ simple path on $G_p$. This restriction on $\mathcal{C}_L$ to satisfy the hardware connectivity graph $(V_p, E_p)$ simplifies the implementation of $M$ without sacrificing access to qubit assignments with higher fidelity: for any fixed $\mathbf{v}_L$ only two of the $n!$ orderings of $\{v_1, \dots v_n\}$ will respect nearest neighbor connectivity on hardware, while the remainder will require additional overhead to implement SWAP networks that will reduce the fidelity of the program and are therefore ignored here. Given the correspondence between $\mathcal{C}_p$ and a simple path described by an array $\mathbf{v} \in V_p^n$, we explicitly define the neighborhood around $\mathcal{C}_p$ as
\begin{equation}
    S(\mathcal{C}_p) = \{\mathbf{v}': 0<|\{v_i'\}_{i=1}^n-\{v_i\}_{i=1}^n|\leq k \} \cup \{R(\mathbf{v}) \}
\end{equation}
where $R$ is the reversal operation sending $[x_1, \dots, x_n] \rightarrow [x_n, \dots, x_1]$,  $|A-B|$ denotes a set difference over elements of $A$ and $B$, and $k\in \mathbb{Z}$ is a tunable parameter that enforces that two neighboring assignments differ by no more than $k$ qubits and edges. By construction every state in the state space of length-$n$ simple paths over $G_p$ may be reached by performing updates to $\mathcal{C}_p$ according to $S(\mathcal{C}_p)$.

In simulations we also considered the more general case of unconstrained logical circuit connectivity thereby extending the state space to include qubit routing or \textit{transpilation}, the task of introducing SWAP networks into a circuit so that circuit entangling operations respect the hardware connectivity. An $n$ qubit circuit without connectivity constraints mapped onto $N$ hardware qubits will result in a state space of size at least $N!/(N-n)!$, greatly complicating the optimization procedure compared to the cases of constrained circuits. The outcome of transiplation varies significantly depending on the initial qubit assignment \cite{holmes_impact_2020} and may involve choosing (deterministically or stochastically) between SWAP networks of equal gate count without any consideration for the device noise. This hinders the ability of an SA algorithm to iteratively explore the full state space of qubit assignments using an ``out-of-the-box'' transpiler (see Appendix~\ref{app:qubit_set_discussion}) and emphasizes the need for noise-aware transpilation. In Appendix~\ref{app:annealer_algorithms} we provide Algorithm~\ref{alg:sa_alltoall} that implicitly defines the neighborhood $S(\mathcal{C}_p)$ around $\mathcal{C}_p$ in the case of unconstrained logical connectivity.

\section{Experiments}\label{sec:experiments}
\subsection{Simulations}

\begin{figure}[ht]
    \centering
    \includegraphics[width=\columnwidth]{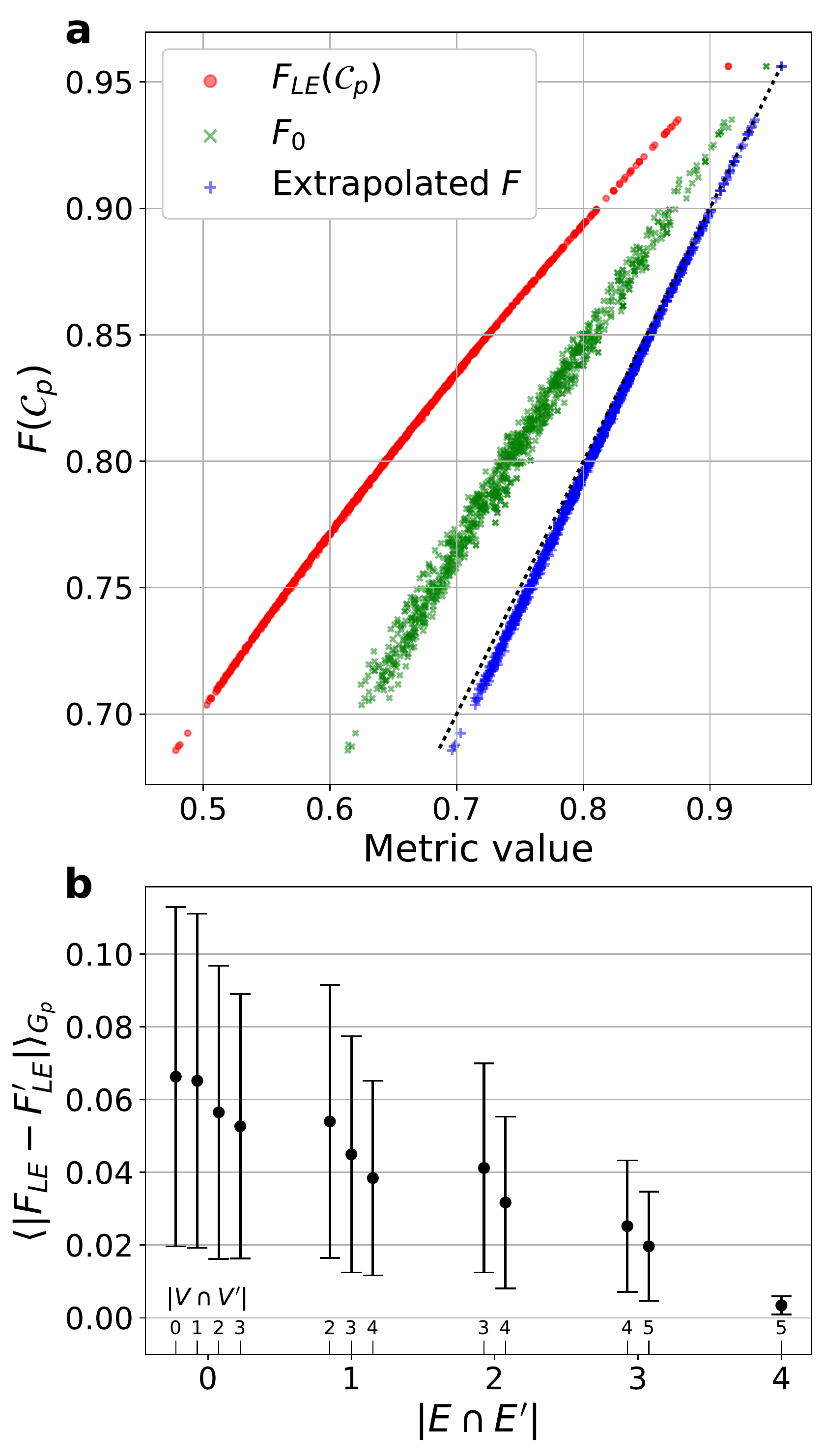}
        \caption{\textbf{a.} Simulated 5-qubit GHZ circuit exposed to a 2-local depolarizing noise model shows a strong monotonic relationship between $\loschmidtf$ and $F$ compared to a weaker relationship between $F_0$ and $F$. Also shown is the extrapolation $F \approx \frac{1}{2}\left(\loschmidtf F_0^{-1} + F_0\right)$ (Eq.~\ref{eq:firstorder_correction}) which closely matches $F$ in the low error limit $F\rightarrow 1$ well outside of the single-error limit (compare to dashed line $y=x$, indicating perfect prediction of $F(\mathcal{C}_p)$). \textbf{b.} Computing the difference in the cost function for every pair of qubit assignments in $G_p$ as a function of the number of shared qubits and edges demonstrates that the cost function is local in $S$. Note that even when all edges and qubits are shared ($k=0$) any linear path and its reversal may differ in the cost function.
        }
    \label{fig:noisemap_1}
\end{figure}

To test our techniques in a controlled setting we constructed a connectivity graph $G_p=(V, E)$ encoding information about the (simulated) fidelities of single and two-qubit gates executed at each vertex and edge in $G_p$. We used the same noise model described in Proposition~\ref{prop:1} consisting of a series of symmetric depolarizing channels $\mathcal{D}_d$ acting on $d$-dimensional density matrices according to
\begin{align}\label{eq:depol}
    \mathcal{D}_d (p): \rho \rightarrow p\frac{I_d}{d} + (1-p) \rho 
\end{align}
which can be simulated on any subsystem of a state $\rho$ using appropriately defined local Kraus operators. After assigning the weight $\epsilon_i$  to each qubit $i \in V$ and $\eta_{ij}$ to each $\{i, j\}\in E$, we construct a noise model by inserting a depolarizing channel $\mathcal{D}_2(\epsilon_i)$ after each single-qubit operation on $i$ and $\mathcal{D}_4(\eta_{ij})$ after each two-qubit operation acting on $\{i, j\}$. Circuit simulations were performed using the \textsc{qiskit} \cite{qiskit} and \textsc{cirq} \cite{cirq} software libraries.

\begin{figure*}[ht]
    \centering
         \includegraphics[width=\textwidth]{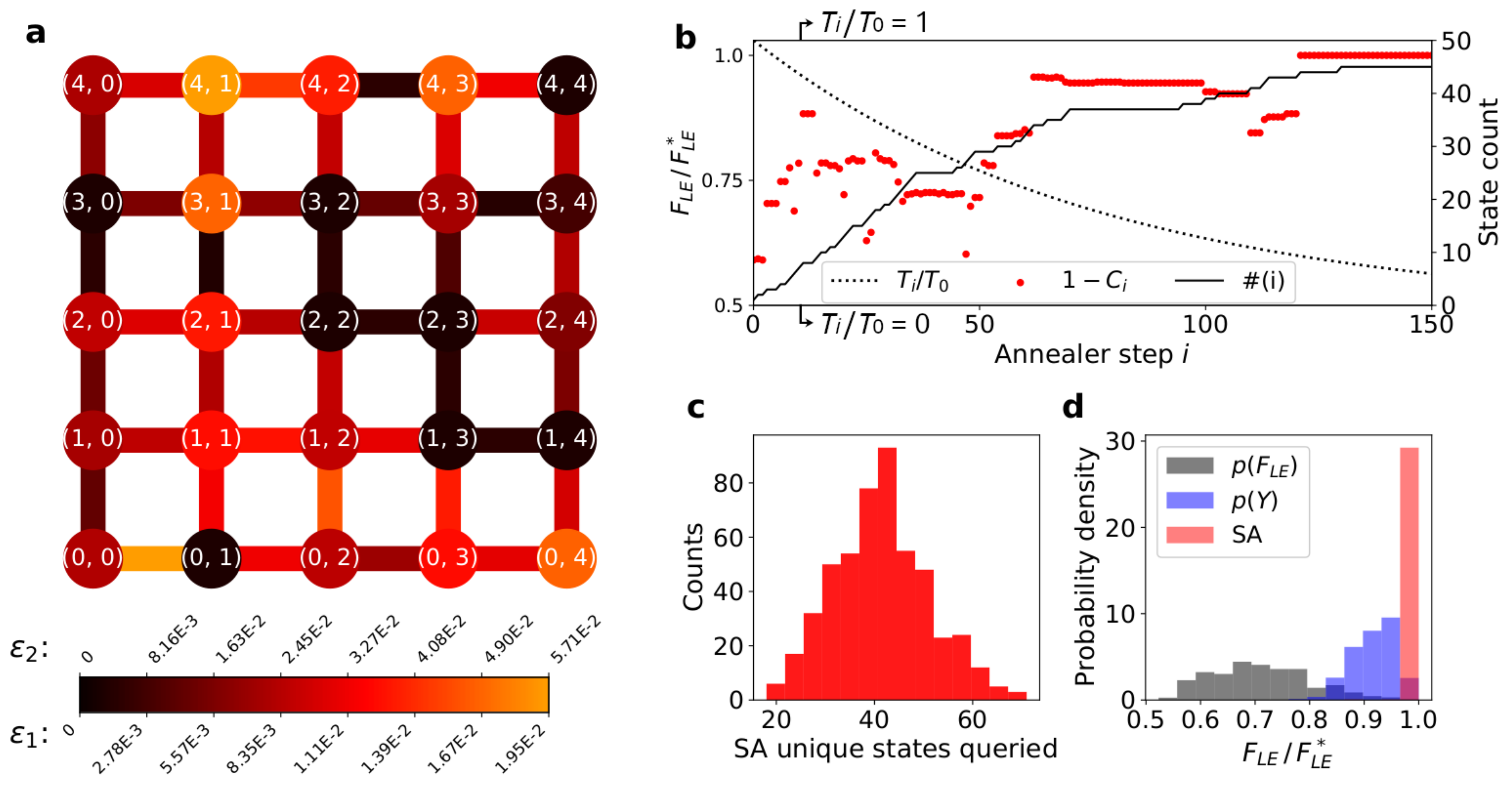}
          \caption{\textbf{a.} Sample simulated noise graph engineered to contain an optimal $5$-qubit assignment used to validate SA qubit assignment in simulation. Each vertex and edge is color coded by the weight corresponding to the depolarization probability for channels acting after each one- and two-qubit gate respectively \textbf{b.} Sample score history for a Simulated Annealer optimization run using simulated noise model on the engineered noisemap, with exponential temperature decay $T_i = T_0 \alpha^i$ ($\alpha = 0.987$, $T_0=0.10$) and neighborhood size $k_{max}=2$ for \edit{150} steps. The score $1 - C_i$ and number of unique states $\#(i)$ queried initially change quickly at high temperatures, before stabilizing as $T$ approaches zero ($S^*$ is often found and maintained early). \edit{The dotted curve for $T_i/T_0$ is plotted on a linear, 0 (bottom of the y-axis) to 1 (top of the y-axis) scale.}  \textbf{c.} The distribution $p(n)$ describing the number of unique assignments sampled by 500 trials of SA is generally smaller than the total number of iterations. \textbf{d.} We compare SA to assignment based on random sampling: Given the set of scores $S$, for 10,000 trials of $n \sim p(n)$ we sample a subset of scores $K \subset S$ with $|K| = n$ and compute $Y = \max ( K )$. The resulting distribution $p(Y)$ is consistently outperformed by SA. Also shown is $P(F_{LE})$, the distribution of scores over all qubit assignments.}
     \label{fig:sim_SA_exp}
\end{figure*}

We first tested the performance of Loschmidt Echoes in simulation by preparing noise graphs with a predetermined optimal qubit assignment. Each noise graph was uniquely determined by its connectivity and the arrays $\boldsymbol{\epsilon}$ and $\boldsymbol{\eta}$. Fig.~\ref{fig:noisemap_1}a shows the results of computing $F_{LE}$ and $F_0(\boldsymbol{\epsilon}, \boldsymbol{\eta})$ for every possible qubit assignment on a $5 \times 5$ simulated grid. We find good agreement between the single-error extrapolation $F \approx \frac{1}{2}\left(\loschmidtf F_0^{-1} + F_0\right)$ derived in Appendix~\ref{app:prop1} to approximate $F$ in the low-error regime, validating our analysis of the Loschmidt Echo.

We also considered the locality of the SA cost function, which determines how much the cost function can change at each iteration of annealing. Given a current qubit assignment $(V, E)$, a neighborhood size of $k$ for the line connectivity state update rule enforces that any new assignment $(V', E')$ satisfies $|V' - V| \leq k$ (for example $k=0$ implies $|E \cap E'| = 4$ and $|V \cap V'| = 5$ for all $(V', E') \subset G_p$). The ability of SA to navigate the space of qubit assignments non-randomly depends on the locality of the cost function defined in Eq.~\ref{eq:cost_def} which we choose to measure using the quantity
\begin{equation}\label{eq:locality}
\langle |F_{LE} - F_{LE}'| \rangle_{G_p} = \underset{\mathcal{C}_p \in S}{\mathbb{E}} \,\, \underset{\mathcal{C}_p' \in S(\mathcal{C}_p)}{\mathbb{E}} |F_{LE} - F_{LE}'|
\end{equation}
Eq.~\ref{eq:locality} describes the typical change in cost moving between assignments for a given definition of neighborhood $S(\mathcal{C}_p)$. Fig.~\ref{fig:noisemap_1}b shows that $\langle |F_{LE} - F_{LE}'| \rangle_{G_p}$ consistently decreases with $k$, indicating that the cost function $C_{LE}$ is local in the space of qubit assignments.

We then tested the performance of SA for qubit assignment using Loschmidt Echoes. Fig.~\ref{fig:sim_SA_exp} demonstrates example performance of SA qubit assignment using $F_{LE}$ as a cost function for a simulated noisy device where the logical and physical connectivity and gatesets were equivalent, $\mathcal{C}_L = \mathcal{C}_p$ for all assignments. Our technique reliably finds high-performing and optimal assignments in noisy simulations. SA consistently outperforms a comparable baseline which keeps the highest performing assignment from a set of $n_s$ random assignments, where $n_s$ is the number of unique assignments attempted by the simulated annealer. We also considered SA for all-to-all logical connectivity circuits which were then transpiled onto nearest neighbor connectivity as a step in the qubit mapping $M$. We found SA to consistently outperformed random qubit assignment, demonstrating that SA for qubit assignment is a viable tool for the assignment of unconstrained logical connecitivity circuits (Appendix~\ref{app:annealer_algorithms}). However, the optimality guarantees for SA using a transpiler are somewhat weaker due to the mismatch between the size of the set of logical circuits and the set of all possible transpiled circuits, emphasizing the need for the development of noise-aware transpilers that can be integrated into a search over qubit assignments.

\subsection{Hardware Experiments}\label{sec:hardware}

We performed experiments using the 23-qubit Rainbow and 52-qubit Weber devices accessed via Google Quantum Cloud Services. Both devices are based on 2D grid nearest-neighbor connectivity (see Fig.~\ref{fig:hardware_diagrams} in Appendix~\ref{app:circuits}). Since readout error significantly biases the estimation of $\loschmidtf$ we implemented readout error correction for all experiments (see Appendix~\ref{app:readout_error}). 

\begin{figure*}[ht]
    \centering
    \includegraphics[width=\textwidth]{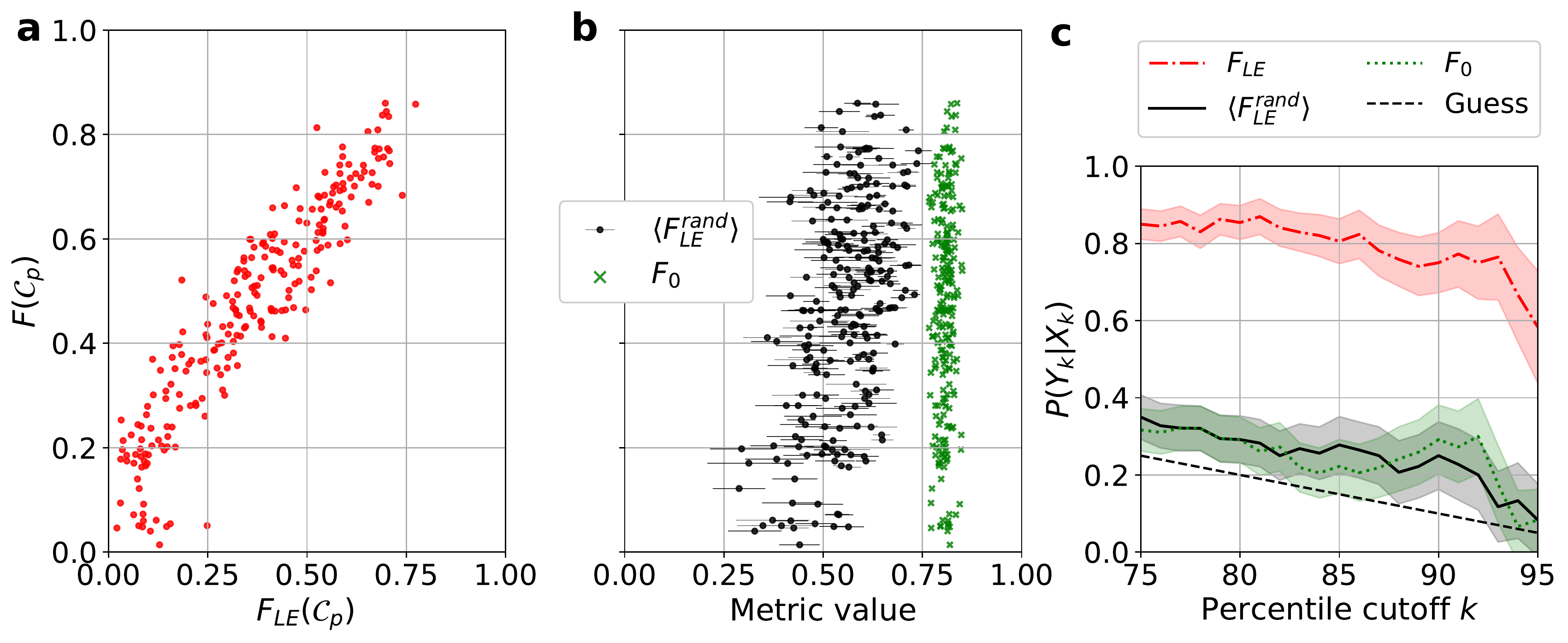}
        \caption{ \textbf{a.} Experimental results comparing $F$ to $F_{LE}$ for an 8-qubit GHZ state demonstrate strong monotonic relationship ($\tau_b = 0.782$) between the two diagnostics.  \textbf{b.} Experiments comparing the average Loschmidt survival over 5 random circuits $\langle F_{LE}^{rand} \rangle$ ($\tau_b = 0.294$) and the metric $F_0:= F_0(\hat{\boldsymbol{\epsilon}}, \hat{\boldsymbol{\eta}})$ ($\tau_b = 0.094$) computed from calibration data $(\hat{\boldsymbol{\epsilon}}, \hat{\boldsymbol{\eta}})$ queried from the quantum processor's most recent calibration (typically 4-8 hours prior to an experiment) show only a weak relationship between either metric and the fidelity $F$. \edit{Error bars for $\langle F_{LE}^{rand}\rangle$ indicate the sample standard deviation computed over 5 random circuits.} \textbf{c.} The diagnostic results can be interpreted by computing the conditional probability of the fidelity $F$ in the $k^{th}$ percentile of fidelities given that an observed metric $X\in \{F_{LE}, \langle F_{LE}^{rand} \rangle, F_0\}$ falls in the $k^{th}$ percentile $X_k$ of its own distribution. Shading indicates standard deviation obtained via bootstrapping. For reference, the dashed line describes uniformly random sampling of $F$ which gives $P(Y_k | X_k) = 1 - k/100$.}
    \label{fig:ghz_results}
\end{figure*}

We first validated the use of the Loschmidt Echo metric $\loschmidtf$ introduced in Sec.~\ref{sec:loschmidt_echo} as a tool for determining the relative performance of qubit assignments. To test the performance of $\loschmidtf$ as a proxy for maximizing the final state fidelity we estimated both $\loschmidtf$ and the fidelity $F$ for a subset of possible assignments on the Rainbow device. We executed three kinds of circuits. The first type of circuit prepares a Greenberger–Horne–Zeilinger (GHZ) state $|\psi_{GHZ}\rangle = \frac{1}{\sqrt{2}}(|0\rangle^{\otimes n} + |1\rangle^{\otimes n})$ using $n=8$ qubits. $F$ was computed using $O(n)$ experiments per qubit assignment using DFE \cite{G_hne_2007} \edit{(see Appendix~\ref{sec:ghz_dfe})}. The second circuit was a SWAP network (SWAPnet): We first prepared an arbitrary single qubit state $|\psi_{SWAPnet}\rangle = \alpha |0\rangle + \beta |1\rangle$ and then moved the state to a different position on the hardware using $8$ sequential pairs of $\sqrt{\text{iSWAP}}$ gates. The fidelity of the SWAPnet can be evaluated in constant time by performing a single-qubit projective measurement onto $|\psi_{SWAPnet}\rangle \langle \psi_{SWAPnet}|$. The final circuit we considered was a Clifford conjugation circuit (or mirror circuit in Refs. \cite{proctor2020measuring,proctor2021scalable}), which has the form
\begin{equation}
U = H^{\otimes n} C^\dagger P C H^{\otimes n}
\end{equation}
where $C \in \text{Cl}(2^n)$ is a Clifford circuit over $n$ qubits, $P = \sigma_{p_1} \otimes \dots \otimes \sigma_{p_n}$ is an $n$-local Pauli string with each $p_k \in \{I, X, Y, Z\}$, and we used $n=8$ for this experiment. While computing $F$ for a Clifford circuit using DFE can incur significant overhead in the general case, here the effect of $U$ simplifies to a Pauli operator acting on a computational basis state. Therefore the output of this circuit is uniquely described by a computational basis bitstring that can be efficiently simulated according to the Gottesman-Knill theorem \cite{gottesman1998heisenberg} such that $F$ can be determined in constant time.

We estimated each of the metrics $\loschmidtf$, $\langle F_{LE}^{rand} \rangle$, and $F_0$ on a subset of qubit assignments satisfying linear connectivity on the Rainbow device, for each choice of circuit described above. 
Fig.~\ref{fig:ghz_results} demonstrates results for diagnostic runs for the GHZ state (corresponding figures for the other circuits can be found in~Appendix~\ref{app:circuits}).
Each fidelity $F$ was computed using DFE using $t=1.5 \times 10^4$ repetitions for each choice of circuit. 
The corresponding values for $\loschmidtf$ for each qubit assignment were computed with variance $\text{Var}(\loschmidtf) = \loschmidtf(1 - \loschmidtf) t^{-1}$ \edit{resulting} in negligible statistical error. Values for $\langle \loschmidtf^{rand} \rangle$ were computed using an average over different random circuits, with $\loschmidtf$ being estimated using $t$ repetitions for each. $F_0(\boldsymbol{\hat{\epsilon}}, \boldsymbol{\hat{\eta}})$ was computed for each circuit assignment using gate counts $n_1$, $n_2$ for the corresponding circuit and taking $\boldsymbol{\hat{\epsilon}}$ to be the average $RB$ error per gate and $\boldsymbol{\hat{\eta}}$ to be the average $\sqrt{\text{iSWAP}}$ XEB error per cycle \cite{isakov2021simulations,google_qcs} (all circuits were implemented using the hardware-native $\sqrt{\text{iSWAP}}$ entangling gate).
\edit{All $F_0$ values were calculated using RB and XEB calibration data estimated in parallel for separate qubits, and we have verified that calculating $F_0$ using non-parallel calibration data has no significant effect on our analysis.}

The goal of using $\loschmidtf$ is to determine the qubit assignment corresponding to the highest value of $F$ by proxy, which will only be possible if there is a strong monotonic relationship between $\loschmidtf$ and $F$. We chose to compute Kendall's $\tau_b$ coefficient \cite{kendall1938new} which captures the concordance between two random variables $X \sim p_X$ and $Y \sim p_Y$ according to
\begin{equation}
\tau_b (X, Y) = \mathbb{E}_{X,Y}\Big[\text{sgn}(x_i - x_j)\,\text{sgn}(y_i - y_j)\Big]
\end{equation}
evaluated over all pairs $(i, j)$ with $i,j=1\dots N$. This coefficient has a simple interpretation in terms of joint behavior of changing $X$ and $Y$. A value $\tau_b=k$ indicates that when the variable $X$ increases (decreases), then the event that $Y$ also increases (decreases) is a factor of $k$ more likely than the event that $Y$ decreases (increases). $\tau_b=0$ indicates no monotonic relationship between variables. Table~\ref{tab:tau_results} computes $\tau_b$ with respect to the fidelity $F$ for each diagnostic on each circuit implemented. In addition, we consider a quantity describing conditional performance of a metric as 
\begin{align}
    P(Y_{k} | X_{k}) &\equiv \frac{P(X > X_k, Y > Y_k )}{P(X > X_k)}
\end{align}
which provides the probability of a signal $Y$ exceeding the $k^{th}$ percentile $Y_k$ of $p_Y$ when an input $X$ is drawn from $p_X$ subject to $X > X_k$, where $X_k$ is the $k^{th}$ percentile of $p_X$. Practically, this quantity predicts the likelihood of a qubit assignment having high fidelity relative to other assignments when the assignment is chosen using only knowledge of the distribution of a diagnostic $X \in \{\loschmidtf, \langle F_{LE}^{rand} \rangle, F_0\}$. Both $\tau_b$ and $P(Y_{k}|X_{k})$  are consistently the highest between $\loschmidtf$ and $F$ compared to other metrics considered, indicating that the Loschmidt Echo is the most reliable proxy when direct measurement of $F$ is unavailable.
\setlength{\extrarowheight}{3pt}
\setlength{\belowrulesep}{0mm}
\begin{table}[ht]
    \centering
        \begin{tabular*}{\columnwidth}{c@{\extracolsep{\fill}}cccc} 
        \toprule
        Circuit & GHZ & Clifford & SWAPnet \\\hline
        $n$ & 8 & 8 & 9 \\
        $m$ & 275 & 350 & 250 \\
        depth & 17 & 25 & 17 \\\hline
        \edit{Diagnostic $X$} & \multicolumn{3}{c}{$\tau_b (X, F)$ $[10^{-2}]$ } \\
        \cmidrule(r){2-4}
        $F_{LE}$ & \textbf{78.2} (1.4) & \textbf{57.0} (2.9) & \textbf{63.5} (2.4) \\
        $\langle F_{LE}^{rand} \rangle$ & 29.4 (3.8) & 52.7 (2.7) & 31.8 (4.0) \\
        $F_0$  & 9.4 (4.4) & 10.4 (4.1) & -10.1 (4.5) \\\hline
        \edit{Diagnostic $X$} & \multicolumn{3}{c}{$P(F_{85}|X_{85})$ $[10^{-2}]$ } \\
        \cmidrule(r){2-4}
        $F_{LE}$ & \textbf{80.5} (5.7) & \textbf{56.4} (6.9) & \textbf{60.0} (9.0) \\
        $\langle F_{LE}^{rand} \rangle$ & 27.8 (7.2) & 53.8 (7.2) & 14.3 (7.5) \\
        $F_0$  & 22.2 (6.8) & 25.6 (6.8) & 8.6 (6.2) \\
        \bottomrule
        \end{tabular*} 
    \caption{Monotonic relationship coefficients $\tau_b$ and  probabilities $P(F_{k}|X_{k})$ for the $k=85^{th}$ percentile computed for each experiment and each diagnostic $X \in \{\loschmidtf, \langle F_{LE}^{rand} \rangle, F_0\}$ evaluated on Rainbow. Standard deviations computed via bootstrapping \cite{efron1994introduction} are provided in parentheses. Each experiment was run for $t=1.5 \times 10^4$ repetitions on each of $m$ distinct, randomly drawn assignments consisting of $n$ qubits composing a simple path sampled from the Rainbow device using networkx \cite{networkx}. For $n=8$ and $n=9$ there are 2984 and 4972 such paths respectively (including reversed paths), highlighting the need for optimization heuristics. In each case, $F_{LE}$ captures a stronger monotonic relationship to $F$ than the other metrics considered, though $\langle F_{LE}^{rand} \rangle$ is an effective method for assessing the performance of other random circuits. }
    \label{tab:tau_results}
\end{table}

We now present the results of SA for qubit selection on a hardware device. For demonstration purposes we consider the case of \textit{offline} qubit assignment, in which the value of $\loschmidtf$ was determined for every possible assignment and then cached in a lookup table to allow for many SA experiments to be repeated. For $n \gtrsim 4$, the number of qubit assignments grows super-exponentially and it becomes infeasible to construct a comprehensive lookup table, in which case SA would need to be performed \textit{online} to optimize the qubit assignment.

\begin{figure}[ht]
    \centering
         \includegraphics[width=\columnwidth]{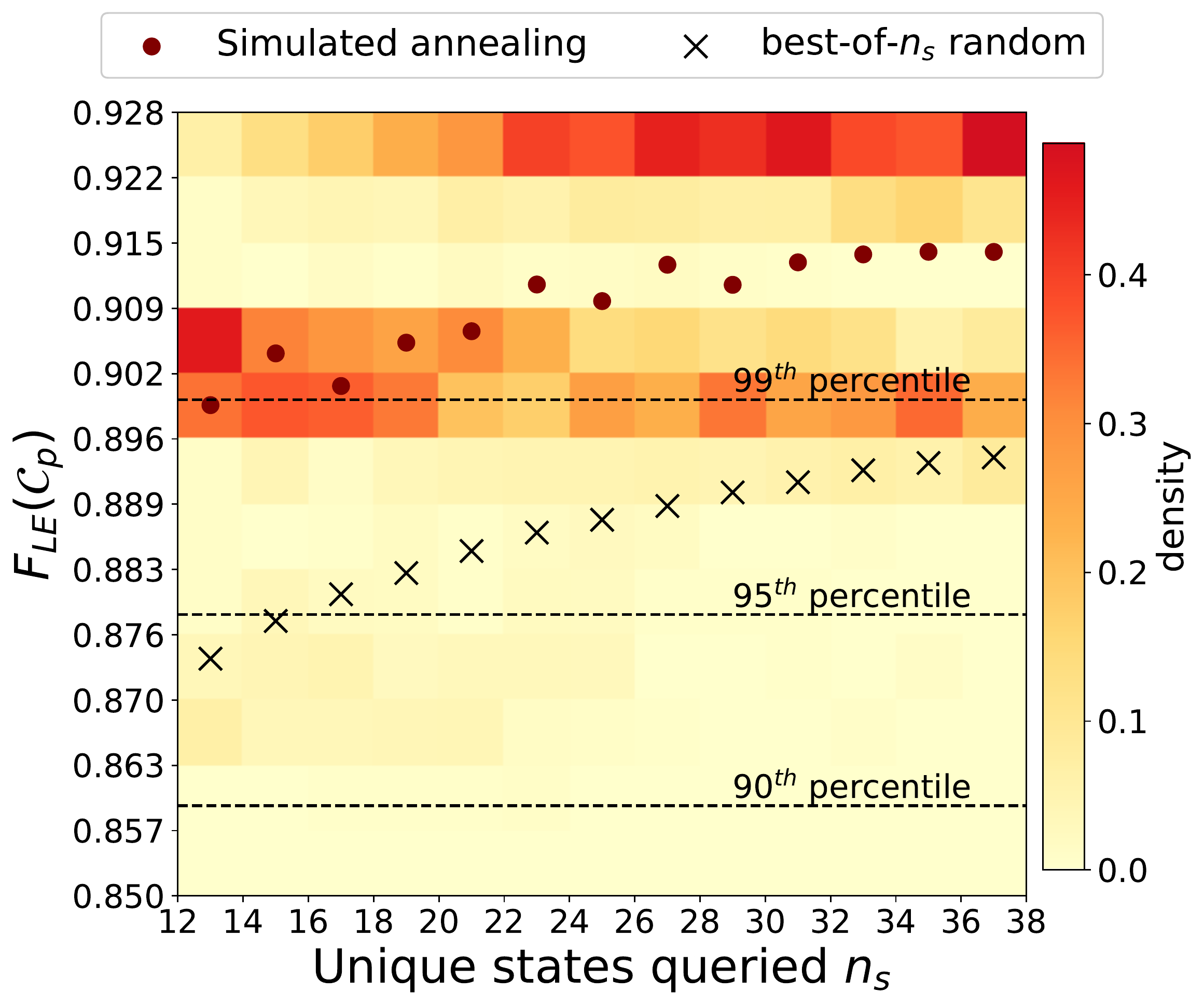}
          \caption{2D histogram of best $\loschmidtf$ values found over 2500 trials of hardware SA using $\loschmidtf$ values computed for $n=4$ GHZ over the complete state space of circuit assignments on Weber. All annealing trials used exponential temperature decay $T_i = T_0 \alpha^i$ ($\alpha = 0.988$, $T_0=0.07$), chosen to balance exploration of the state space with suppressing $n_s$. Columns are renormalized since the typical number of unique states queried $n_s$ concentrates around 23 for this choice of SA hyperparameters. SA optimization frequently finds the best possible value $\loschmidtf$ from the state space (contained in the maximium $y$-bin shown) but the $n_s$ necessary to do so varies. Given the poor predictive capabilities of device diagnostic data for GHZ state fidelity, we compare this method to a random sampling scheme that consists of drawing $n_s$ circuit assignments without replacement (10000 trials per $n_s$) and keeping the assignment corresponding to the maximum $\loschmidtf$ observed. SA consistently outperforms random sampling for the $n=4$ GHZ dataset. Averaged over all choices of $n_s$ the final $F_{LE}$ achieved by SA outperformed the random sampling scheme by 2.8\%. } 
     \label{fig:hardware_SA_exp}
\end{figure}

To test the performance of Simulated Annealing for qubit assignment on hardware, we prepared a $4$-qubit GHZ state restricted to logical line connectivity for each of the 1116 possible length-$4$ simple paths on the 53-qubit Weber grid layout. We evaluated  both orderings of qubit assignments over each simple path, with the expectation that each ordering may produce a different fidelity. Roughly $27\%$ of assignments were rejected on account of exceed our maximum threshold for 0.15 qubit visibility (see Appendix~\ref{app:readout_error}), with SA being performed over the remaining dataset of 808 simple paths. Fig.~\ref{fig:hardware_SA_exp} shows the results of SA for qubit selection using $\loschmidtf$. SA optimization runs are divided according to the number of unique states $n_s$ for which $\loschmidtf$ would need to be computed on hardware during an online run. SA frequently found the qubit assignment corresponding to the optimal $\loschmidtf^*$ using $n_s$ as small as 30, fewer than $5\%$ of the assignments available in the dataset. Since prior knowledge of $F_0$ computed from diagnostic data could not be relied on to find a high scoring qubit assignment, we chose best-of-$n_s$ random sampling as a competitive baseline optimization technique on the dataset. We found that SA consistently outperformed this scheme for all choices of $n_s$. 

 We also performed SA for qubit assignment on the 23 qubit Rainbow device for an $n=3$ Quantum Fourier Transform (QFT) circuit. Each circuit was executed with a specific input basis state $|j\rangle,\, j \in [0, 2^n - 1]$ so that DFE could be performed using $O(1)$ experiments. Due to the much smaller state space (only 148 possible length-3 simple paths) we found that random assignment was competitive with SA (see Appendix~\ref{sec:qft}). This highlights the importance of considering circuit size and device layout when considering the qubit assignment problem.

\section{Conclusions}

Even with the availability of detailed error rates for a quantum device, the task of assigning a logical circuit to a subset of hardware qubits that maximize the fidelity of the prepared state can be extremely difficult. In this work we have demonstrated  instances of circuits for which the Loschmidt Echo provides an effective stand-in for the fidelity function in qubit assignment, and have provided theoretical justification for this relationship in the small-error limit. Conversely, we have shown that even excellent knowledge of device error rates (e.g. computed using Randomized Benchmarking) can be insufficient to predict the performance of a circuit mapped onto hardware. \edit{Furthermore, we provided analytical arguments that the advantage of $F_{LE}$ over $F_0$ for determining relative performance would persist even if $F_0$ could be estimated to higher precision using scalable techniques (e.g. Refs.~\cite{Harper_2020,flammia_2021})}. We have also showed that Simulated Annealing can be used to find performant qubit assignments using significantly fewer resources than a complete exploration of possible qubit assignments would require. 

Our theoretical arguments have relied on a low-error limit which limits this analysis to circuits with either few qubits or low gate depth. However our experimental results indicate that a strong connection between the Loschmidt Echo and the state fidelity function may still hold in a regime of higher errors. Future work may determine that such a relationship exists for a broader class of noise models or outside of the perturbative regime.

In this work we have evaluated the use of a Loschmidt Echo diagnostic on a specific selection of circuits employing only readout error mitigation, and so the optimal qubit assignments determined in this work may not remain optimal when employing platform-specific error mitigation. In particular, control errors which may go undetected by (see Appendix~\ref{app:counterexamples}) can be mitigated using Floquet Calibration \cite{Neill2021}, thereby strengthening the relationship between $F_{LE}$ and $F$. 
Therefore such device-specific calibration procedures should be combined with qubit assignment based on $\loschmidtf$ to maximize the final assignment performance. 
\edit{Other error mitigation strategies may be found, for example in \cite{Kandala2019,doi:10.1126/sciadv.aaw5686}.}

However, as the number of qubit assignments grows exponentially with respect to the number of available qubits on the device and the number of qubits in the circuit, it becomes essential to minimize the experimental overhead associated with device calibration and computation of $F_{LE}$ for each assignment.
Future work will be necessary to determine the appropriate balance of implementing error mitigation so that $F_{LE}$ remains faithful to $F$ while minimizing overhead such that a significant number of qubit assignments can be probed during the optimization procedure. 

While we have focused on maximizing the fidelity of a state prepared by a specific unitary $U$, our method can be readily adapted to analyze more general situations. For instance, a Hilbert-Schmidt test \cite{Khatri_2019} which computes $\Tr(U^\dagger V)$ for unitaries $U, V$ could be used to compute  $\Tr (\noisyop{U^\dagger}\circ \noisyopon{U}{\zerodm} )$ as a proxy for the process fidelity \cite{PhysRevA.60.1888,NIELSEN2002249}, $F(U, \noisyop{U}) = \int \mathcal{F}(U\psi U^\dagger, \mathcal{E}_U (\psi) ) d\psi$ . However such a modification will introduce significant two-qubit gate overhead, for which our weak-error analysis of the Loschmidt echo technique is less applicable.

\edit{\subsection{Data and Code availability}
Code and data to reproduce the analyses in this manuscript are available online \cite{coderelease}.
}

\section{Acknowledgements}

We thank the Google Quantum AI team for access to their quantum computing hardware and for all of the support they provided for experiments.
We thank Alan Ho, Ryan LaRose, Xiao Mi, and Matthew Harrigan for thoughtful conversations about our experiments.
We thank Khalil Guy and Andy Sun for contributing to an earlier version of this project. 
This work was partially supported by the DOE/HEP QuantISED program grant ``HEP Machine Learning and Optimization Go Quantum,'' identification number 0000240323 and the DOE/HEP QuantISED program grant ``QCCFP-QMLQCF Consortium,'' identification number DE-SC0019219.
Access to the Google QPU was supported under the Fermilab LDRD project FNAL-LDRD-2018-025 ``Towards a Quantum Computing Science Center at Fermilab.''
EP is partially supported through Achim Kempf's Google Faculty Award. This manuscript has been authored by Fermi Research Alliance, LLC under Contract No. DE-AC02-07CH11359 with the U.S. Department of Energy, Office of Science, Office of High Energy Physics.

\appendix

\section{Optimal qubit assignments with respect to Loschmidt Echo and state fidelity}
\label{app:prop1}

\subsection{\edit{Local depolarizing noise} \delete{Proof of Proposition 1}}\label{app:depolarizing_noise_channel}
A significant claim of this work is that under some assumptions on error rates and circuit sizes, performing qubit assignment with respect to the Loschmidt Echo metric  $\loschmidtf$ will provide reasonable outcomes compared to performing qubit assignment with respect to the fidelity $F$ of a state prepared by applying a noisy implementation of $U$ to an initial state $|0\rangle$. As described in the main text, we use the definitions

\begin{align}
    \loschmidtfon{\mathcal{C}_p} &= \Tr \left( \zerodm  \noisyop{U^\dagger} \circ \noisyopon{U}{\zerodm} \right)\\
    F(\mathcal{C}_p) &= \langle 0 |U^\dagger  \noisyopon{U}{\zerodm} U |0 \rangle 
\end{align}

\noindent
where $\noisyop{V}$ is a noisy implementation of the unitary $V$ which implicitly depends on the parameters of the circuit $\mathcal{C}_p$. \edit{We wish to compare} the optimal qubit set $S_{LE}^*$ that maximizes $\loschmidtf$ with respect to a hardware connectivity graph (weighted by error rates) to the qubit set $S^*$ that maximizes $F$:

\begin{align}
    S_{LE}^* &= \{ S': S'\subset S, S' = \argmax_{\mathcal{C}_p \in S} \loschmidtfon{\mathcal{C}_p '} \} \\
    S^* &= \{ S': S'\subset S, S' = \argmax_{\mathcal{C}_p \in S} F(\mathcal{C}_p ') \}  
\end{align}

\noindent
where $S$ describes the complete state space of hardware circuits that realize the operation $U$ in the absence of noise, $S = \{\mathcal{C}_p: U(\mathcal{C}_p) = U\}$. For notational convenience we will typically write $F = F(\mathcal{C}_p)$, $\loschmidtf = \loschmidtfon{\mathcal{C}_p}$. Each metric will be evaluated with respect to a unitary $U = \prod_{i=1}^m U_i$ 
acting on an input state prepared to $|0\rangle$. We assume without loss of generality that each $U_i$ is up to two-local. We will consider a discrete noise model that can be described by insertion of CPTP maps within an existing circuit, with each map being parameterized by error rates that depend only on the choice of qubit or qubit pair involved in the execution of a gate. We emphasize that an assumption about the device noise is necessary for any comparison between $\loschmidtf$ and $F$ since there exists a family of simple noise models for which these two quantities are completely independent, as demonstrated in Appendix~\ref{app:counterexamples}.

We therefore choose a noise model that inserts error channels for each single- and two-qubit gate occuring in a circuit. Fix a device connectivity graph as $G_p = (E_p, V_p)$ and a circuit $\mathcal{C}_p$ acting over $n$ qubits chosen from $N=|V_p|$ qubits with entangling operations subject to the edges provided in $E_p$. The unitary $U$ implemented by $\mathcal{C}_p$ can be described by a sequence of single qubit gates acting on individual nodes $p$ and two-qubit gates acting on edges $\{p, q\}\in E_p$. The circuit contains a total of $n_p$ gates applied to each node $i$ (where $n_p=0$ if $\mathcal{C}_p$ does not apply any single qubit gates to $i$) and $n_{ij}$ gates applied to each pair of qubits $\{p, q\}$.

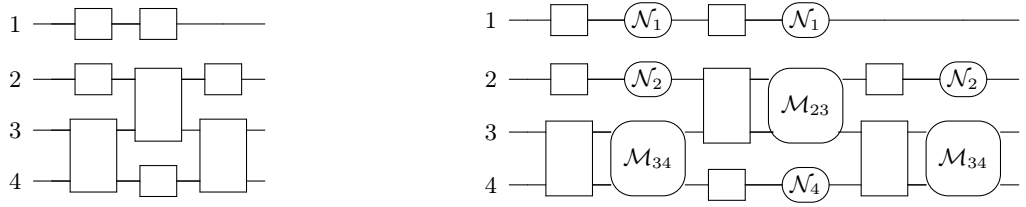
\begin{figure*}
    \begin{minipage}{0.45\textwidth}
        \[ 
    \Qcircuit @R=1em @C=0.75em {
         \\
         &\lstick{\text{1}}& \qw&\gate{\phantom{H}}               \qw&\gate{\phantom{H}}                  \qw&      \qw&\qw\\
         &\lstick{\text{2}}& \qw&\gate{\phantom{H}}                      \qw&\multigate{1}{\phantom{H}}  \qw&\gate{\phantom{H}}   \qw&\qw\\
         &\lstick{\text{3}}& \qw&\multigate{1}{\phantom{H}}  \qw&\ghost{\phantom{H}}        \qw&\multigate{1}{\phantom{H}} &\qw\\
         &\lstick{\text{4}}& \qw&\ghost{\phantom{H}}         \qw&\gate{\phantom{H}}              \qw&\ghost{\phantom{H}}   \qw&\qw\\
         \\
        }
        \]
    \end{minipage}
    \begin{minipage}{0.45\textwidth}
        \[ 
    \Qcircuit @R=1em @C=0.75em {
         \\
         &\lstick{\text{1}}& \qw&\gate{\phantom{H}}           &\measure{\mathcal{N}_1}             \qw&\gate{\phantom{H}}           &\measure{\mathcal{N}_{1}}             \qw&                       \qw&      \qw&\qw\\
         &\lstick{\text{2}}& \qw&\gate{\phantom{H}}             &\measure{\mathcal{N}_{2}}             \qw&\multigate{1}{\phantom{H}}&\multimeasure{1}{\mathcal{M}_{{23}}} \qw&\gate{\phantom{H}}            &\measure{\mathcal{N}_{2}}  \qw&\qw\\
         &\lstick{\text{3}}& \qw&\multigate{1}{\phantom{H}}   &\multimeasure{1}{\mathcal{M}_{{34}}} \qw&\ghost{\phantom{H}}       &\ghost{\mathcal{N}_{{23}}}        \qw&\multigate{1}{\phantom{H}}    &\multimeasure{1}{\mathcal{M}_{{34}}}&\qw\\
         &\lstick{\text{4}}& \qw&\ghost{\phantom{H}}          &\ghost{\mathcal{M}_{{34}}}        \qw&\gate{\phantom{H}}        &\measure{\mathcal{N}_{4}}             \qw&\ghost{\phantom{H}}         &\ghost{\mathcal{M}_{{34}}}  \qw&\qw\\
         \\
        }
        \]
    \end{minipage}
    \caption{The two-local noise model here appends channels to each gate depending on which vertex or edge the preceding unitary was applied to. Rounded boxes indicate completely-positive trace preserving operators that are not unitary in general (so that the diagram on the right represents maps acting $n$-qubit density matrices). If we consider $\mathcal{N}$ to be a channel that applies a non-identity operator with probability $\epsilon$  then the state $|\psi_{(i, j)}\rangle$ describes the output of the noisy operator in the case where a non-identity operator was drawn from the Kraus representation of $\mathcal{N}$. While a more general model could consider $\mathcal{N}_p$ and $\mathcal{M}_{pq}$ to be dependent on the gate prior to their insertion point, we do not consider this level of detail in this work.}
\label{fig:noise_v2}
\end{figure*}

In the noisy implementation $\noisyop{U}$ each single-qubit gate acting on qubit $p\in V_p$ is followed by some fixed, CPTP single-qubit map $\mathcal{N}_{p}$ also acting on $p$, and each two-qubit gate on $\{p, q\}$ is followed by a two-qubit channel $\mathcal{M}_{pq}$ acting on $\{p, q\}$. For convenience in the derivation, we will define a single-qubit error rate $\epsilon_p$ such that the Kraus operator representation of $\mathcal{N}_p$ is given as

\begin{align}
    \mathcal{N}_p (\rho) &= \sum_{A \in K_p} A \rho A^\dagger \\
    K_p &= \{\sqrt{1-\epsilon_p} \I\} \cup \{E_1, \dots E_{r_p}\} 
    \\\sum_{k=1}^{r_p} E_k^\dagger E_k &= \I\epsilon_p
\end{align}

\noindent
so that $\epsilon_p$ is the probability that $\mathcal{N}_p$ applies a nontrivial error operation. 
We define $\eta_{pq}$ to similarly be the probability that $\mathcal{M}_{pq}$ acts with a nontrivial error operation on edge $\{p, q\}$. In this sense, $\epsilon_p$ and $\eta_{pq}$ describe weights in the hardware connectivity graph $G_p$. Fig.~\ref{fig:noise_v2} illustrates an example construction of this noise model. 

From $\mathcal{N}$ we define another map

\begin{align}\label{eq:Nsub}
    \mathcal{N}_p^{sub} (\rho) := \sum_{k=1}^{r_p} \epsilon_p^{-1} E_k \rho E_k
\end{align}

\noindent
which contains the (nontrivial) errors that occur when $\mathcal{N}$ is applied to some state $\rho$. Assuming that the operation $U_k$ is $i$-th operation applied to qubit $p$ we can then define the (renormalized) mixture over the nontrivial Kraus operators for the $i^{th}$ channel $\mathcal{N}_p$ acting on qubit $p$ during execution of $U$

\begin{align}
    \rho_{(p, i)} &:= \mathcal{U}_m \circ \cdots \circ\mathcal{U}_{k+1} \circ \mathcal{N}_p^{sub} \circ \mathcal{U}_k \circ \cdots \circ \mathcal{U}_1(\zerodm)
\end{align}

\noindent
where we have used the shorthand $\mathcal{U}(\rho) = U \rho U^\dagger$. The  state $\rho_{(pq, i)}$ can be defined analogously as the result of inserting $\mathcal{M}_{pq}$ after the $i$-th operation applied to edge $\{p, q\}$, yielding the mixture over the nontrivial Kraus operators for the $i^{th}$ channel $\mathcal{M}_{pq}$ acting on edge $\{p,q\}$, and $|\psi\rangle = U|0\rangle$. This permits a compact analytical expression for the effect of $\noisyop{U}$ and $\noisyop{U^\dagger}$ for this form of noise model. The state prepared by $\noisyopon{U}{\zerodm}$ is given by

\begin{align}
    \noisyopon{U}{\zerodm} = F_0\Bigl(&|\psi\rangle\langle \psi| + \sum_{p\in V_p} \sum_{i=1}^{n_p} \frac{\epsilon_p}{1-\epsilon_p} \rho_{(p, i)} \nonumber
    \\&+ \sum_{\{p,q\}\in E_p} \sum_{i=1}^{n_{pq}} \frac{\eta_{pq}}{1-\eta_{pq}} \rho_{(pq, i)} + \cdots \Bigr) \label{eq:forwards_noise}
\end{align}

\noindent
where we have omitted terms corresponding to two or more errors and substituted the expression

\begin{equation}
    F_0 = \prod_{p\in V_p} ( 1-\epsilon_p)^{n_p} \prod_{\{p,q\}\in E_p} ( 1-\eta_{pq})^{n_{pq}}
\end{equation}

\noindent
originally presented in Eq.~\ref{eq:F0} of the main text. The appearance of $F_0$ here motivates its interpretation as a zero-error contribution to the fidelity $F$. The remainder of this derivation will consider contributions only up to the single error limit (which will generally still involve high-order terms in $\epsilon$ and $\eta$). The fidelity of $U$ can be computed directly as 

\begin{align}\nonumber
    F &= F_0 \Bigl(1 + \sum_{p\in V_p} \sum_{i=1}^{n_p} \frac{\epsilon_p}{1-\epsilon_p}  \langle \psi |\rho_{(p, i)} | \psi \rangle 
    \\ \label{eq:twolocal_f} &+ \sum_{\{p,q\}\in E_p} \sum_{i=1}^{n_{pq}} \frac{\eta_{pq}}{1-\eta_{pq}} \langle\psi |\rho_{(pq, i)}|\psi\rangle+ \cdots \Bigr)
    \\&= F_0 + F_1 + \cdots
\end{align}

\noindent

Defining the quantity that describes drawing a nontrivial error for the $i^{th}$ instance of $\mathcal{N}_p$ acting on the reverse circuit:

\begin{align}
    \mathcal{E}_{U^\dagger}^{(p,i)}\left( |\psi\rangle\langle\psi | \right) &:= \mathcal{U}_1^\dagger \circ \cdots \circ \mathcal{N}_p^{sub} \circ \mathcal{U}_{k}^\dagger \circ \mathcal{U}_{k+1} \circ \cdots \circ \mathcal{U}_m^\dagger(|\psi\rangle\langle\psi |)
\end{align}

\noindent
and defining $\mathcal{E}_{U^\dagger}^{(pq,i)}$ analogously for $\mathcal{M}_{pq}$, we can similarly compute $\loschmidtf$ by applying $\noisyop{U^\dagger}$ to Eq.~\ref{eq:forwards_noise} to arrive at

\begin{widetext}
\begin{align}\nonumber
    \loschmidtf &= F_0^2 \left(1 + \sum_{p\in V_p} \sum_{i=1}^{n_p} \frac{\epsilon_p}{1-\epsilon_p} \left( \langle \psi |\rho_{(p, i)} | \psi \rangle +\langle 0 |\mathcal{E}_{U^\dagger}^{(p,i)} \left( |\psi\rangle\langle\psi | \right) |0\rangle\right) \right. 
    \\\label{eq:twolocal_fle} &\qquad\qquad \left. + \sum_{\{p,q\}\in E_p} \sum_{i=1}^{n_{pq}} \frac{\eta_{pq}}{1-\eta_{pq}} \left(  \langle \psi |\rho_{(pq, i)} | \psi \rangle + \langle 0 |\mathcal{E}_{U^\dagger}^{(pq,i)} \left( |\psi\rangle\langle\psi | \right) |0\rangle \right)+ \cdots \right)
\end{align}
\end{widetext}

While there is evidence that $\loschmidtf$ contains some of the information about the errors involved in the computation of $F$, this relationship is still too general to draw any specific comparisons. So to prove Proposition~\ref{prop:1} relating $\loschmidtf$ and $F$, we instantiate a noise model for which both local channels $\mathcal{N}_p$ and $\mathcal{M}_j$ are symmetric depolarizing channels (Eq.~\ref{eq:depol}) over one and two qubits respectively:

\begin{align}
    \mathcal{N}_p \rightarrow \mathcal{D}_2(\epsilon_p), \qquad  \mathcal{M}_{pq} \rightarrow \mathcal{D}_4(\eta_{pq}) 
\end{align}

This simple model involving local depolarization is sometimes used in the analysis of random circuits \cite{Magesan_2011,Boixo_2018} and will be sufficient to demonstrate interesting behavior in the more general problem of qubit assignment. Each depolarizing operation may be commuted with the preceding unitary gate without affecting the behavior of the global state. Specifically, for the case of single-qubit depolarization where

\begin{align}
    \mathcal{N}_{p}' &= \Bigl(\underbrace{\I_2 \otimes \cdots \otimes \I_2 }_{i-1} \otimes\, \mathcal{D}_2(\epsilon_p) \otimes  \underbrace{\I_2 \otimes \cdots\otimes  \I_2 }_{n-i}\Bigr) \\
    \mathcal{U}_i' &= \Bigl(\underbrace{\I_2 \otimes \cdots \otimes \I_2 }_{i-1} \otimes\, \mathcal{U}_i \otimes  \underbrace{\I_2 \otimes \cdots\otimes  \I_2 }_{n-i}\Bigr)
\end{align}

\noindent
then for each single-qubit gate and channel we have that

\begin{align}
    \mathcal{N}_k'\circ  \mathcal{U}_k'\circ |\psi\rangle\langle \psi | =  \mathcal{U}_k' \circ \mathcal{N}_k'\circ |\psi\rangle\langle \psi |
\end{align}

\noindent
and a similar relation holds for each two-qubit gate acted on by $\mathcal{D}_4$. From the freedom in the ordering of each error and each gate occuring in the noisy adjoint circuit $\noisyop{U^\dagger}$ it follows that 

\begin{align}\label{eq:reverse_same}
    \langle 0 |\mathcal{E}_{U^\dagger}^{(p,i)} \left( |\psi\rangle\langle\psi | \right) |0\rangle &=  \langle \psi |\rho_{(p, i)} | \psi \rangle\\
    \langle 0 |\mathcal{E}_{U^\dagger}^{(pq,i)} \left( |\psi\rangle\langle\psi | \right) |0\rangle &=   \langle \psi |\rho_{(pq, i)} | \psi \rangle
\end{align}

Rescaling each $\epsilon \rightarrow \frac{3\epsilon}{4}$ and $\eta \rightarrow \frac{15\eta}{16}$ to account for a trivial pauli string in the Kraus representation of $\mathcal{D}_d$ and substituting Eq.~\ref{eq:reverse_same} into Eq.~\ref{eq:twolocal_fle} yields 

\begin{align}\nonumber
    \loschmidtf &\rightarrow  F_0^2 \Bigl(1+ 2\Bigl[\sum_{p, i}\frac{\epsilon_p}{1-\epsilon_p}  \langle \psi |\rho_{(p, i)} | \psi \rangle 
    \\&+  \sum_{\{p,q\}, i} \frac{\eta_{pq}}{1-\eta_{pq}}  \langle \psi |\rho_{(pq, i)} | \psi \rangle\Bigr] + \cdots \Bigr)\label{eq:depol_loschmidtf_v2} \\
    &= F_0 \left(F_0 + 2 F_1 + \cdots \right)
\end{align}

We therefore arrive at a relationship between $\loschmidtf$ and $F$ in the single-error limit assuming local depolarizing noise. The final step to arrive at Proposition~\ref{prop:1} is to take the small error rate limit. We recall that $\loschmidtf$, $F$, and $F_0$ are implicitly dependent on a hardware circuit $\mathcal{C}_p=(\textbf{g}, \textbf{v}_L, \textbf{e}_L)$ containing complete information about which qubits and pairs of qubits are acted on by any gate in the gate sequence $\mathbf{g}$. We let $\mathbf{n}$ to be the length-$|V_p|$ vector of gate counts applied to each logical qubit $p \in V_L$ with $(\mathbf{n})_p = n_p = |\{k: 1 \leq k \leq m',\, (\textbf{v}_L)_k = p\}|$, and $\mathbf{m}$ to be the length-$|E_p|$ vector of gate counts applied to each available edge $\{p, q\}\in E_p$ with $(\mathbf{m})_{pq} = n_{pq} = |\{k: 1 \leq k \leq m',\, (\textbf{e}_L)_k = \{p, q\} \}|$. We similarly define $(\boldsymbol{\epsilon})_p = \epsilon_p$ and $(\boldsymbol{\eta})_{pq} = \eta_{pq}$ to contain the corresponding single- and two-qubit error rates. For some fixed ordering of $\boldsymbol{\epsilon}$ and $\boldsymbol{\eta}$, the circuit assignment problem (without introducing SWAP overhead) can then be understood as the procedure of ordering the elements of $\mathbf{n}$ and $\mathbf{m}$ under the constraint that the resulting circuit corresponds to the unitary $U$ and respects the hardware device connectivity.

We now study this optimization by taking the limit that $(n_p \epsilon_p) \ll 1$ and $(n_{pq} \eta_{pq}) \ll 1$ for all $p \in V_p, \, \{p,q\} \in E_p$, and arrive at the perturbative approximation 

\begin{align}
    F_0 &\approx 1 - \mathbf{n} \cdot \mathbf{\epsilon} - \mathbf{m} \cdot \mathbf{\eta}
\end{align}

To repeat this procedure for $F$ and $F_{LE}$, we define

\begin{align}
    (\mathbf{f})_p &= \sum_{i=1}^{n_p} \langle \psi |\rho_{(p, i)} | \psi \rangle, \qquad p \in V_p\\
    (\mathbf{g})_{pq} &= \sum_{p=1}^{n_{pq}} \langle \psi |\rho_{(pq, i)} | \psi \rangle, \qquad \{p, q\} \in E_p
\end{align}

The arrays $\mathbf{f}$ and $\mathbf{g}$ are explicitly dependent on $\mathbf{n}$ and $\mathbf{m}$ respectively, and are therefore also dependent on the choice of qubit assignment. Noting that from Eq.~\ref{eq:Nsub}  we then have that $|(\mathbf{f})_p| \leq n_p$, and similarly $| (\mathbf{g})_p)| \leq n_{pq}$. This guarantees that terms of the form $(\epsilon_p (\mathbf{f})_p)^2$ vanish at least as quickly as $(\epsilon_p n_p)^2$ and therefore permits the perturbative expansion

\begin{align}
    F_1 &\approx \mathbf{f} \cdot \boldsymbol{\epsilon} + \mathbf{g} \cdot \boldsymbol{\eta} \\ \label{eq:FU_approx}
    F &\approx 1 + (\mathbf{f} - \mathbf{n} )\cdot \boldsymbol{\epsilon} + (\mathbf{g} -\mathbf{m} )  \cdot \boldsymbol{\eta}     \\
    F_{LE} &\approx 1 + 2(\mathbf{f} - \mathbf{n} )\cdot \boldsymbol{\epsilon} + 2(\mathbf{g} -\mathbf{m} )  \cdot \boldsymbol{\eta}   
\end{align}

Then under the conditions we imposed, it is clear that under \textit{any} choice of qubit assignment constraints we have 

\begin{equation}\label{eq:Q_QLE}
    S^* = \argmax_{\mathbf{f}, \mathbf{g}, \mathbf{n}, \mathbf{m}} F = \argmax_{\mathbf{f}, \mathbf{g}, \mathbf{n}, \mathbf{m}} \loschmidtf = S_{LE}^*
\end{equation}

\noindent
and therefore under any specific constraints that each choice of $\mathbf{n}, \mathbf{m}$ respects hardware connectivity and realizes the unitary $U$, $F$ and $\loschmidtf$ similarly share a maximum.
\edit{This demonstrates a specific case of Proposition~\ref{prop:1} for the local depolarizing noise model.}
\delete{This completes the proof of Proposition~\ref{prop:1}.}

A number of remarks are necessary to put this result into context. First, the result of Eq.~\ref{eq:Q_QLE} would not necessarily hold without the assumption of local depolarizing noise, reemphasizing that this result works only for a restricted family of noise models. Notably, this kind of noise model can be enforced on a given hardware circuit using randomized compiling techniques \cite{Wallman_2016}, by which the insertion of some redundant single-qubit operations has the effect of converting systematic errors into stochastic depolarizing errors. Furthermore, this result was attained under a restrictive choice of perturbative limit in which the typical number of errors $(\epsilon n)$ is very small (though typically a similar assumption is necessary for fault tolerance required by most forms of quantum error correction). We also note that the terms of the form $\mathbf{f} \cdot \boldsymbol{\epsilon}$ may contribute significantly less to the magnitude of $F$ and $\loschmidtf$ than $\mathbf{n} \cdot \boldsymbol{\epsilon}$, depending on how the intermediate states in the implementation of $U$ are affected by the various Pauli operators imposed by $\mathcal{D}$.

However, even under these restrictions we have that an optimization based on $F_0$ will not necessarily arrive at $S^*$. Even in the case where an experimentalist has perfect knowledge of the depolarizing error rates in a device (computed using RB for example), more detailed knowledge about the effects of the errors (contained in the terms $\mathbf{f}$ and $\mathbf{g}$) is necessary to probe for the optimal qubit assignment. This shortcoming can be demonstrated with a simple example where we suppose $G_N$ has exactly $n-1$ noiseless qubits to choose from (with the remainder being acted on by maximally depolarizing channels) and $U$ is the circuit preparing the state 

\begin{equation}
    \frac{1}{\sqrt{2}}\left( |0^{n-1}\rangle + |1^{n-1}\rangle\right) |0\rangle
\end{equation}

\noindent
In this case, $F_0$ is flat over the space of qubit assignments that contain the $n-1$ noiseless qubits, but $F$ is maximized only when $n^{th}$ logical qubit is assigned to a noisy device qubit since depolarizing any of the qubits in the entangled subsystem of the state will completely eliminate the coherence of the entangled state. 

The truncated expansions for $F$, $\loschmidtf$, and $F_0$ can be validated via numerical simulation. Comparing Eq.~\ref{eq:depol_loschmidtf_v2} to Eq.~\ref{eq:twolocal_fle} gives a single-error approximation for $F$ as

\begin{equation}\label{eq:firstorder_correction}
    F = \frac{1}{2}\left(\frac{\loschmidtf}{F_0} + F_0\right) + \text{(multiple-error terms)}
\end{equation}

\noindent
which allows one to approximate $F$ having knowledge of only $\loschmidtf$ and $F_0$ given this specific noise model.

This relation suggests that it may be possible to estimate $F$ via extrapolation techniques if one can reliably compute $F_0$, for instance using randomized circuits constructed using the same number of gates acting on each qubit and edge as $U$. As this extrapolation assumes an underlying noise model consisting of depolarizing noise, an accurate characterization of performance using the relationship of Eq.~\ref{eq:firstorder_correction} might require techniques such as randomized compiling \cite{Wallman_2016}.

\subsection{Alternative noise models}\label{app:other_noise}

We now discuss the optimization behavior of $F$, $F_0$, and $\loschmidtf$ in the context of other noise models. Appendix~\ref{app:global_depol} will describe a simple noise model for which all three metrics share an optimal qubit assignment, demonstrating that under relaxed conditions knowledge of $F_0$ is sufficient to optimize a qubit assignment. Appendix~\ref{app:counterexamples} will describe a noise model for which $F$ and $\loschmidtf$ are entirely uncorrelated. Appendix~\ref{app:markovian} will extend the results of Appendix~\ref{app:depolarizing_noise_channel} to a family of general Markovian noise models in the weak error limit.

\subsubsection{Global depolarizing noise}\label{app:global_depol}

In contrast to Appendix~\ref{app:depolarizing_noise_channel}, here we study a noise model for which the optimal qubit assignment according to $F_0$, $F_{LE}$, and $F$ is identical. We consider global depolarizing noise acting on the system with probability proportional to the error rate of the most recently executed gate. While such a model is highly unrealistic for most choices of $U$, it is well motivated in the case where $U$ is a random circuit drawn uniformly with respect to the Haar measure on $U(2^n)$. Without loss of generality we may decompose $U$ as $U = \prod_{i=1}^m U_i$ where each $U_i\in G_p$ is a member of the hardware gate set for simplicity. Each gate $U_i$ is understood to act on the vertices and edges $(\mathbf{i}, e_i)$. Then for each $U_i$ we apply a depolarizing channel (Eq.~\ref{eq:depol} with $d=2^n$) acting over all $n$ qubits so that the noisy implementation of $U$ acting on $\zerodm$ is given by

\begin{equation}
    \mathcal{E}_U(\zerodm) = \mathcal{D}_{m} \circ \mathcal{U}_m \circ \cdots \circ \mathcal{D}_{1} \circ \mathcal{U}_1(\zerodm)
\end{equation}

\noindent
where the map $\mathcal{U}_i: \rho \rightarrow U_i \rho U_i^\dagger$ is adopted for convenience. Each channel $\mathcal{D}_i$ has a depolarizing probability given by

\begin{equation}\label{eq:depol_param}
    \epsilon_p = \begin{cases}
    w_V(i) \qquad \text{if } e_i = \emptyset \\
    w_E(e_i) \qquad \text{otherwise}
    \end{cases}
\end{equation}

\noindent
where $w_V$ and $w_E$ denote the weights for a vertex or edge respectively according to a fixed noise graph $G_N$. We can now directly compute $F$ and $\loschmidtf$. Starting from an initial state $\zerodm$ the result of applying the noisy process $U_\mathcal{E}$ can be computed recursively as

\begin{align}
    \mathcal{E}_U(\zerodm) &= \prod_{k=1}^m (1-\epsilon_k) \rho_m + \left(1 -  \prod_{k=1}^m (1-\epsilon_k)\right)\frac{\mathbb{I}}{d} \\\label{eq:rho_m}
    &= F_0 \rho_m + \left(1 -  F_0\right)\frac{\mathbb{I}}{d}
\end{align}

\noindent
where we have substituted $F_0$ defined in Eq.~\ref{eq:F0} with $\eta_{pq}=0$:

\begin{align}
    F_0 &= \prod_{k=1}^m (1-\epsilon_k) = \prod_{k: v_k\in V} (1-\epsilon_k)^{n_k}
\end{align}

\noindent
where $n_k$ counts the multiplicity of each qubit or edge occurring in the instruction set (so that $\sum_k n_k = m$) and $|V|=n$. This is a slight generalization of the result derived in \cite{du2020learnability}.
Assuming that each operation $U_i$ and $U_i^\dagger$ are depolarized by the same amount, we apply $\mathcal{E}_{U^\dagger}$ and then measure the probability of recovering the all zeros bitstring to be

\begin{align}\label{eq:depol_loschmidtf}
    \loschmidtf &=\frac{d-1}{d} F_0^2 + \frac{1}{d}
\end{align}

\noindent
whereas the fidelity of the state prepared by $\mathcal{E}_U$ using Eq.~\ref{eq:rho_m} as 

\begin{equation}\label{eq:depol_truef}
F = \frac{d-1}{d} F_0 + \frac{1}{d}
\end{equation}

Then, defining  $ S^* = \argmax_{S\subseteq V_p} F_0$ and using the monotonicity of $x^2$ for $x>0$ we immediately find

\begin{equation}
    S^* = \argmax_{S\subseteq V_p}  F = \argmax_{S\subseteq V_p}  \loschmidtf = S_{LE}^*
\end{equation}

\noindent
which is the desired result. Furthermore, the fact that $S^*$ was defined as the qubit assignment maximizing $F_0$ means that for this noise model, the optimal qubit assignment can be found by maximizing any one of $F$, $\loschmidtf$, or $F_0$. This result is expected since the noise model does not incorporate any information about the structure of $U$ (only the number of gates acting on each qubit at some point in time), and so the assumption of a digital error model holds and $F_0$ is an effective proxy for the circuit fidelity. This result further highlights that in the case of sufficiently random circuits $F_0$ may be an effective optimizer over qubit assignments. 

In the above analyses we have defined $F_0$ in terms of depolarizing channel parameters $\epsilon_k$ and $\eta_{jk}$, even though though it is not immediately clear that this is the correct parameter to optimize over. In contrast, a depolarizing model can be viewed in terms of a Pauli error $e_p$ by defining the Kraus operators as \cite{arute2019quantum}

\begin{align}\label{eq:depol_pauli}
    E_\ell &= \sqrt{\frac{e_p}{2^n - 1}} P_{\ell}, \qquad \ell \neq 0
    \\E_0 &= \sqrt{1 - e_p} \,\mathbb{I}
   \\ P_\ell &= \sigma_{\ell_1} \otimes \cdots \otimes \sigma_{\ell_n}
\end{align}

\noindent
where $\sigma_\ell$ is a Pauli matrix with $\ell \in \{0,1,2,3\}^n$.
Comparing this action of this channel to a global depolarizing channel with probability $p$ (Eq.~\ref{eq:depol} for $d=2^n$) gives the relationship

\begin{equation}
    e_p = p (1 - 4^{-n})
\end{equation}

Therefore computation of $F_0$ from experimentally measured diagnostic data is subject to ambiguities arising from the interpretation of error rates. The quantity $F_0(\hat{\boldsymbol{\epsilon}}, \hat{\boldsymbol{\eta}})$ is equally valid when its arguments represent individual gate infidelities (which are combined in a multiplicative fashion into a total circuit fidelity by $F_0$) or when its arguments represent probabilities for depolarization after each gate. The parameters in these two error models are related by a constant scalar, but it is straightforward to show that the ranking of qubit assignments with respect to $F_0(c\hat{\boldsymbol{\epsilon}}, c\hat{\boldsymbol{\eta}})$ will not necessarily be preserved with respect to a rescaling factor $c$ applied to the error rates. Define the function

\begin{align}
    f(x, c) &= \prod_{i=1}^{d} (1 - cx_i)^{n_i}
\end{align}

\noindent
for any $x \in \mathbb{R}_+^d$, $n \in \mathbb{Z}_+^d$ and $0 \leq c \leq 1$ (the upper bound on $c$ may be loosened as long as $cx_i \leq 1 \,\forall \, i$). Then letting the vectors $\mathbf{\epsilon}$ and $\mathbf{\epsilon}'$ be the per-qubit error rates for assignments $S$ and $S'$ respectively, it is easy to show that there exist choices $\mathbf{\epsilon}$, $\mathbf{\epsilon}'$, $c$, and $c'$ such that $f(\mathbf{\epsilon}, c) > f(\mathbf{\epsilon'}, c)$ but $f(\mathbf{\epsilon}, c') < f(\mathbf{\epsilon'}, c')$. Then letting $c=1,\, c' = (1-4^{-n})$, one can generally find a distribution of errors such that the optimal assignment differs depending on whether $F_0$ is computed with respect to the Pauli error rate $e_p$ or the depolarization parameter $p$. This further complicates the problem of optimal assignments using $F_0$ computed from calibration data, as there is no a priori choice of scaling that best suits the assignment problem with respect to a given unitary $U$.


\subsubsection{Gate-dependent unitary noise}\label{app:counterexamples}

In contrast to Appendix~\ref{app:depolarizing_noise_channel}, is straightforward to construct a noise model for which $S^* \neq S_{LE}^*$. Consider if $\mathcal{E}_U$ is strictly unitary, such that $\mathcal{E}_U(\rho) = \tilde{U} \rho \tilde{U}^\dagger$ and 

\begin{align}
\tilde{U} &= \prod_{i=1}^m W_i U_i
\end{align}

If the (unitary) noise affecting the reverse circuit is its own adjoint,

\begin{equation}
    \mathcal{E}_{U^\dagger}(\rho) = \tilde{U}^\dagger \rho \tilde{U}
\end{equation}

\noindent
then $\loschmidtf = 1$ for any choice of $\{ W_i\}$ while $\mathcal{F}(U)$ can take on arbitrary values in $[0, 1]$. Such a model might be relevant to some forms of unitary control error and highlights the need for some contextual understanding of the noise processes involved in a specific hardware when using the technique we introduced in this work. For instance in \cite{mi2021information} it was observed that similar unitary error occurred for both $(\text{iSWAP})^{1/2}$ and $(\text{iSWAP})^{-1/2}$ which would result in accumulated errors reflected in $F$ being possibly cancelled out during the computation of $\loschmidtf$.

\subsection{Perturbative limit with Markovian noise}\label{app:markovian}

\newcommand{\dg}{^{\dagger}}
\newcommand{\ds}[2]{\frac{\mathrm{d} {#1} }{\mathrm{d} {#2}} }
\newcommand{\oA}{A}
\newcommand{\oH}{H}
\newcommand{\orho}{\rho}
\newcommand{\sD}{\mathcal{D}}
\newcommand{\sDj}[1]{\sD_{#1}}
\newcommand{\sH}{\mathcal{H}}
\newcommand{\sHj}[1]{\sH_{#1}}
\newcommand{\dtj}[1]{\Delta t_{#1}}

\newcommand{\sA}{\mathcal{A}}
\newcommand{\sE}{\mathcal{E}}
\newcommand{\sEj}[1]{\sE_{#1}}

\newcommand{\be}{\begin{equation}}
\newcommand{\ee}{\end{equation}}
\newcommand{\lf}{\left}
\newcommand{\rt}{\right}
\newcommand{\nn}{\nonumber}

\newcommand{\Trb}[1]{\Tr \lf[ #1 \rt]}
\newcommand{\sandwich}[3]{\lf< #1 \lf| #3 \rt| #2 \rt>}

We consider a more general case such that the infidelity can be modeled as a result of weak Markovian noises and time-independent control errors.
The time evolution of the qubit density matrix $\orho$ is governed by a Lindblad master equation \cite{breuer2002},
\begin{equation}
	\ds{}{t} \orho=  -i [\oH, \orho]+ \sum_j \frac{\gamma_j}{2} ( 2 \oA_j \orho \oA_j\dg - \oA_j\dg \oA_j \orho - \oA_j\dg \oA_j \orho),
\end{equation}
where $\oH$ is the qubit Hamiltonian representing the qubit gates and control errors, and the Markovian noise channels are represented by the jump operator $\oA_j$ and the decay rate $\gamma_j$.
To make the discussion more convenient, we introduce the superoperators which map an operator to another operator. Similar to operators, additions, multiplications, Hermitian transpose and inverse are well-defined over these superoperators.
In particular, we can define
$\sH \orho = -i [\oH, \orho]$ and
$\sD \orho = \sum_j \frac{\gamma_j}{2} ( 2 \oA_j \orho \oA_j\dg - \oA_j\dg \oA_j \orho - \oA_j\dg \oA_j \orho)$
to express the master equation as
\begin{equation}
	\label{eq:app_master_equation}
	\ds{}{t} \orho= \sH \orho + \sD \orho.
\end{equation}

A quantum program can be modeled as a series of $m$ gates sequentially applied to the QPUs.
Each of the gate is represented by a noiseless propagator $U_j = e^{-i \sHj{j;0} \dtj{j} }$ generated by the gate Hamiltonian $\sHj{j;0}$ where $j=1, 2, \cdots, m$ and $\dtj{j}$ is the gate duration time.
In the presence of noise, the propagator has to be modified according to the master equation \eqref{eq:app_master_equation} which has a formal solution $\rho(t) = e^{\sH t + \sD  t } \rho(t=0)$.
The noisy propagator $\noisyop{U}$ of a quantum program $U$ can hence be written as
\begin{equation}
	\noisyop{U} = \prod_{j=1}^{m} \noisyop{U_j} = \prod_{j=1}^{m} e^{\sHj{j} \dtj{j} + \sDj{j} \dtj{j}},
\end{equation}
where $\noisyop{U_j}$ is the noisy propagator of the $j$-th qubit gate.
The Hamiltonian $\sHj{j} = \sHj{j;0} + \sHj{e;U_j}$ consists of the gate Hamiltonian $\sHj{j;0}$ and the static control error Hamiltonian $\sHj{e;U_j}$ associated with the gate $U_j$.
In principle, the control error could be different for the gate $U_j$ and its conjugate $U_j\dg$ and we have
\begin{align}
\noisyop{U_j} = & e^{\sHj{j;0} \dtj{j} + \sHj{e;U_j} \dtj{j}   + \sDj{j} \dtj{j}},
\\
\noisyop{U_j\dg} = & e^{-\sHj{j;0} \dtj{j} + \sHj{e;U_j\dg} \dtj{j}   + \sDj{j} \dtj{j}}.
\end{align}
We put a negative sign in front of $\sHj{j;0}$ for $\noisyop{U_j\dg}$ since the conjugate gate $U_j\dg$ can be generated by $- \sHj{j;0}$. In the above equations, we also assume that the Markovian noise channels are the same for $U_j$ and $U_j\dg$.
To simplify the following discussion, we define
\begin{equation}
	\sHj{j;e\pm} = \frac{1}{2} (\sHj{e;U_j} \pm \sHj{e;U_j\dg}).
\end{equation}

The fidelity $F$ and the Loschmidt Echo metric $\loschmidtf$ as in the previous discussions are given by
\begin{align}
	F &=
	\langle 0 |U^\dagger  \noisyopon{U}{\zerodm} U |0 \rangle 
	=
	 \Tr \lf( \orho_T  \orho_1  \rt) ,
	\\
	\loschmidtf
	&=
	\Tr \left( \zerodm  \noisyop{U^\dagger} \circ \noisyopon{U}{\zerodm} \right)
	= \Tr \left( \orho_0  \orho_2 \right).
\end{align}
Here, $\orho_T = |\psi\rangle\langle\psi |$ is the target pure state prepared by the noiseless $U$, $\orho_0 = \zerodm $ is the initial pure state, $\orho_1$ is the state prepared by the noisy propagator $\noisyop{U}$, and $\orho_2$ is the state prepared by the noisy propagator $\noisyop{U}$ followed by the noisy inverse propagator $\noisyop{U\dg}$.
We can also express these states in terms of the superoperators,
\begin{align}
	\orho_T = &
	\prod_{j=1}^{m} e^{\sHj{j;0} \dtj{j}} \orho_0
	,
	\\
	\orho_1 = &
	\prod_{j=1}^{m} e^{\sHj{j;0} \dtj{j} + \sHj{j;e+} \dtj{j} + \sHj{j;e-} \dtj{j}  + \sDj{j} \dtj{j}} \orho_0
	,
	\\
	\orho_2 = &
	\prod_{j=m}^{1}
	e^{-\sHj{j;0} \dtj{j} + \sHj{j;e+} \dtj{j} - \sHj{j;e-} \dtj{j}  + \sDj{j} \dtj{j}}
	\orho_1
	.
\end{align}
We assume that the gate duration times $\dtj{j}$ are small to separate the noiseless evolution and the error terms such that
\begin{align}
	\orho_1 = &
	\prod_{j=1}^{m}
	e^{\sHj{j;e+} \dtj{j} + \sHj{j;e-} \dtj{j}  + \sDj{j} \dtj{j}}
	e^{\sHj{j;0} \dtj{j}}
	\orho_0
	\nn\\
	& + \mathcal{O} \lf(\dtj{j}^2 \rt)
	,
	\\
	\orho_2 = &
	\prod_{j=m}^{1}
	e^{-\sHj{j;0} \dtj{j}}
	e^{\sHj{j;e+} \dtj{j} - \sHj{j;e-} \dtj{j}  + \sDj{j} \dtj{j}}
	\orho_1
	\nn\\
	& + \mathcal{O} \lf(\dtj{j}^2 \rt)
	.
\end{align}
This decomposition is valid as long as the gate duration time is much smaller than the characteristic time scale of the errors.
We then introduce $\epsilon$ as an order-counting parameter such that $\sHj{j;e+} \rightarrow \epsilon \sHj{j;e+}$, $\sHj{j;ei} \rightarrow \epsilon \sHj{j;e-}$ and $\sDj{j} \rightarrow \epsilon \sDj{j}$.
Expanding the exponentials in the above equations thus gives us
\begin{align}
	\orho_1 = &
	\prod_{j=1}^{m}
	\lf[
	1 + \epsilon \dtj{j} \lf( \sHj{j;e+}  +  \sHj{j;e-} + \sDj{j} \rt)
	\rt]
	e^{\sHj{j;0} \dtj{j}}
	\rho_0
	\nn\\
	&
	+ \mathcal{O} \lf(\epsilon^2, \dtj{j}^2 \rt)
	,
	\\
	\orho_2 = &
	\prod_{j=m}^{1}
	e^{-\sHj{j;0} \dtj{j}} 
	\lf[
	1 + \epsilon \dtj{j} \lf(\sHj{j;e+} - \sHj{j;e-} + \sDj{j}\rt) 
	\rt]
	\orho_1
	\nn\\
	&
	+ \mathcal{O} \lf(\epsilon^2, \dtj{j}^2 \rt)
	.
\end{align}

In the perturbative limit, the error terms are small such that we can take the leading terms in the above equation to compute $F$ and $\loschmidtf$. We then have
\begin{widetext}
\begin{align}
	F = &
	1 +
	\epsilon
	\sum_{k=1}^{m}
	\Trb{
		\orho_0
		\lf(\prod_{j=k}^{1} e^{-\sHj{j;0} \dtj{j}}\rt)
		\dtj{k} \lf(\sHj{k;e+} + \sHj{k;e-} + \sDj{k}\rt) 
		\lf(\prod_{j=1}^{k} e^{\sHj{j;0} \dtj{j}}\rt)
		\orho_0}
	,
\\
	\loschmidtf = &
	1 +
	2 \epsilon
	\sum_{k=1}^{m}
	\Trb{
		\orho_0
		\lf(\prod_{j=k}^{1} e^{-\sHj{j;0} \dtj{j}}\rt)
		\dtj{k} \lf(\sHj{k;e+}  + \sDj{k}\rt) 
		\lf(\prod_{j=1}^{k} e^{\sHj{j;0} \dtj{j}}\rt)
		\orho_0}
\end{align}
\end{widetext}
Comparison these two equations, we arrive at the following relation
\be
\label{eq:pert_F_FLE}
F - 1 = \frac{\loschmidtf - 1}{2} + E_{e-} + \mathcal{O}(\epsilon^2)
\ee
where
\begin{align}
    E_{e-} = & \epsilon
    \sum_{k=1}^{m}
    \mathrm{Tr}
    \Bigg[
    	\orho_0
    	\lf(\prod_{j=k}^{1} e^{-\sHj{j;0} \dtj{j}}\rt)
    	\dtj{k}  \sHj{k;e-}
    \nn\\
    	&  \quad\quad\quad\quad \times
    	\lf(\prod_{j=1}^{k} e^{\sHj{j;0} \dtj{j}}\rt)
    	\orho_0
    \Bigg]
\end{align}
is the control error dependent on the pulse sequence realizations of $U$ and $U\dg$.

If the control error of any gate is the same to that of its inverse implementation, $E_{e-}$ vanishes. We can rearrange equation \eqref{eq:pert_F_FLE} for $\loschmidtf$ results in the relationship
\be
\loschmidtf = 2F - 1 +  \mathcal{O}(\epsilon^2).
\ee
This means that in the weak error limit, $\loschmidtf$ is linearly proportional to $F$. Then in general, the ranking of $\loschmidtf$ and the ranking of $F$ are the same for noise models satisfying the following conditions:
\begin{enumerate}
	\item The environmental noise is Markovian such that the time evolution of the system density matrix can be described by a Lindblad master equation \cite{breuer2002};
	\item The control error is the same for any gate $U$ and its inverse $U^{\dagger}$;
	\item The system is in the weak-error limit;
	\item The gate duration time is much smaller than the characteristic time scale of the errors.
\end{enumerate}
\edit{
In other words, given these conditions, the optimal assignments with respect to $F$ and $\loschmidtf$ are the same, i.e.\
\begin{equation}
	S^* =  S_{LE}^*,
\end{equation}
which completes the proof of Proposition~\ref{prop:1}.
}
One can show that the depolarizing noise models discussed in Appendix~\ref{app:depolarizing_noise_channel} and Appendix~\ref{app:other_noise} are special cases of theses conditions with no control errors and the error rates given  by $\gamma_j \dtj{j}$ for the specific depolarizing channels.
One can also check that the gate-dependent unitary noise defined in Appendix~\ref{app:counterexamples} violates the second condition and thus $F_{LE}$ no longer serves as a good approximation of $F$.
	
\section{Readout error mitigation}\label{app:readout_error}

We discuss the effect of readout error on $\loschmidtfon{U}$ and techniques for efficient readout error mitigation. In this case let $\mathcal{E}_U\rightarrow \mathcal{U}$ and  $\mathcal{E}_{U^\dagger}\rightarrow \mathcal{U}^\dagger$, but replace the perfect measurement operation with a noisy one. A noisy computational basis measurement can be represented by the POVM $\{P_j\}$ with $j=0\dots 2^n-1$ with each element $P_j$ given by

\begin{equation}\label{eq:measurement_noise}
    P_j = \sum_{k} p_{jk} |k\rangle \langle k|
\end{equation}

\noindent
where the restriction that $\sum_{j} p_{jk} = 1$ ensures that $\sum_{j} P_j = I$. The choice $P_j = |j\rangle\langle j|$ recovers a noiseless computational basis measurement. For a fixed $j$, each $p_{jk}$ in  Eq.~\ref{eq:measurement_noise} can be interpreted as the probability of measuring computational basis state $|k\rangle$ but then recording that result as $|j\rangle$. Substituting $P_0$ for $\zerodm$ in Eq.~\ref{eq:roundtrip_fid} of the main text we find that

\begin{align}
    \loschmidtfon{U} &= \text{Tr}\left(P_0 U U^\dagger\zerodm U^\dagger U \right)
    \\&= \text{Tr}\left(P_0 \zerodm\right)
    \\&= p_{00}
\end{align}

\noindent
which shows that this fidelity estimate will only depend on the probability that the bitstring $0^n$ is measured correctly by the noisy measurement, and the qubit set $\tilde{Q}^*$ resulting from optimization of $\loschmidtfon{U}$ in the presence of readout error will be biased towards maximizing $p_{00}$. Clearly this does not fully characterize the effect of readout error on a measurement applied to $\noisyopon{U}{\zerodm}$, and so $\tilde{Q}^*$ will generally differ from a qubit set $Q^*$ that minimizes the effect of readout error after preparing, say, $\mathcal{E}_U(\zerodm)$. In an experimental setting this bias can be removed by applying readout error mitigation which can efficiently recover the probability of the all-zeros bitstring from the output of a quantum device \cite{peters2021perturbative}.

In exploratory experiments we determined that readout error correction assuming uncorrelated bitflip errors for each qubit closely matched the performance of readout error correction assuming correlated bitflips. \edit{We performed readout error correction for all experiments by first determining the conditional error probability $p_{01}=p(0 |1)$ of observing ``0'' given a pre-measurement state $|1\rangle$ and $p_{10}=p(1 |0)$ of observing ``1'' given a pre-measurement state $|0\rangle$ for every qubit. We then ran  experiments to empirically estimate the probability $p'(j)$ of observing each bitstring $j=0, \dots, 2^n$ with readout error. Since $p'(j)$ represents a joint probability over single bit outcomes, we then estimated the true probability $p(j)$ of measuring bitstring $j$ using the tensor product of a set of linear maps used to correct each pair of single-bit marginal probabilities $p'(0)$ and $p'(1)$, each having the form}

\begin{equation}
    \begin{pmatrix}
    p(0) \\ p(1) 
    \end{pmatrix} = \begin{pmatrix}
    p(0 |0) & p(0 |1) \\
    p(1 |0) & p(1 |1)
    \end{pmatrix}^{-1}\begin{pmatrix}
    p'(0) \\ p'(1) 
    \end{pmatrix}.
\end{equation}

\noindent
The error propagated by a linear equation of the form $\mathbf{p} = Q^{-1} \mathbf{p}'$ increases with increasing condition number $\kappa(Q) = \norm{Q} \norm{Q^{-1}}$, and therefore greater readout error rates $p(x | \neg x)$ will lead to a breakdown in the performance of readout error correction. We therefore rejected qubit assignments for which $\max ( p(0 |1), p(1 |0) ) > 0.15 $, typically fewer than $5 \%$ of assignments on Rainbow but $\sim 27\%$ assignments on Weber. We found that the relationship between $F$ and $\loschmidtf$ was consistently strengthened by performing readout error correction. This is partly due to the tendency of asymmetric readout error rates (i.e. $p(0 |1) > p(1 |0)$) to increase $F_{LE}$ (since the experimentalist infers greater survival likelihood from a larger population of observed $0^n$ bitstrings) while at the same time decreasing $F$ (in which case the effect of mostly random readout error with high probability increases the distance of the observed bitstring distribution to the bitstring distribution associated with the output of DFE).

\section{Mapping logical circuits to physical circuits}\label{app:qubit_set_discussion}

We present a more detailed description of the qubit assignment problem. We consider a map acting on a logical circuit that returns a hardware circuit performing the same operation as the logical circuit:

\begin{equation}
M: (\mathcal{G}_L^m, V_L^m, E_L^m) \rightarrow (\mathcal{G}_p^{m'}, V_p^{m'}, E_p^{m'})
\end{equation}

\noindent
and assume that $\mathcal{C}_L \in (\mathcal{G}_L^m, V_L^m, E_L^m)$ is uses exactly $n$ unique qubits taken from $V_L$. We further subject $M$ to the constraint that $U(\mathcal{C}_p)|0\rangle$ executed in a noiseless setting will prepare the same state as executing the logical circuit. We should allow for the use of additional operations or qubits in the execution of $\mathcal{C}_p$ that do not affect the unitary $U(\mathcal{C}_L)$, and so we let $\mathcal{C}_p \in (\mathcal{G}_p^{m'}, V_p^{m'}, E_p^{m'})$ use exactly $N$ qubits taken from $V_p$ and constrain the map $M$ such that 

\begin{equation}
    U(\mathcal{C}_L)|0\rangle\langle0|U(\mathcal{C}_L)^\dagger = \Tr_{A}\left(U(\mathcal{C}_p)|0^N\rangle \langle 0^N | U(\mathcal{C}_p)^\dagger \right)
\end{equation}

\noindent
for some register $A$ consisting of at most $(N-n)$ qubits. We will consider implementing a specific unitary $U$. To allow for a layer of abstraction between the logical circuit and physical implementations, we fix a map $M$ and consider the sets of permissible logical and physical circuits to be

\begin{align}
    \mathbf{C}_L &= \{\mathcal{C}_L: \mathcal{C}_L \in (\mathcal{G}_L^m, V_L^m, E_L^m), \, U(\mathcal{C}_L) = U\} \\
    \mathbf{C}_p &= \nonumber\{\mathcal{C}_p: \mathcal{C}_p \in (\mathcal{G}_p^{m'}, V_p^{m'}, E_p^{m'}), \\ & \qquad \qquad (\exists \mathcal{C}_L \in (\mathcal{G}_L^m, V_L^m, E_L^m):  \mathcal{C}_p = M(\mathcal{C}_L)) \}   
\end{align}

To implement Simulated Annealing we fix the set of possible states of the annealer to be $S = \mathbf{C}_p$, and the algorithm proceeds by iteratively evaluating the cost function for elements drawn from $S$ and updating to another element of $S$ accordingly. We now discuss how the definition of $\mathbf{C}_p$ with respect to a mapping $M$ complicates the implementation of SA.

In our hardware experiments we chose $M$ to be the identity mapping and set $V_p = V_L$, $E_p = E_L$, and $\mathcal{G}_p = \mathcal{G}_L$. As a result, $\mathbf{C}_p = \mathbf{C}_L$. This effectively separates compilation of logical operations into physical operations from the qubit assignment. If we define an update rule such that every element of $\mathbf{C}_L$ may be reached in a finite number of update steps from any other element, then every element of $S$ may similarly be reached after a finite number of steps and so the algorithm is guaranteed to converge by the conditions presented in Ref. \cite{bertsimas1993simulated}. 

However in general, compilation of logical circuits to hardware will modify the contents of a logical circuit $\mathcal{C}_L$ such that $\mathbf{C}_L \neq \mathbf{C}_p$. Alternatively, error mitigation strategies such as dynamical decoupling \cite{viola1999dynamical,PhysRev.80.580} or zero noise extrapolation \cite{2017_zne_a,2017_zne_b} may introduce additional operations to $\mathcal{C}_p$ that do not affect the realized operation $U(\mathcal{C}_p)$. Similarly, ``wait gates'' which implement the identity operation with fidelity $\leq 1$ on hardware can be used to probe $T_1$ and $T_2^*$.  Under these considerations, a reasonable choice of $M$ might allow any of the following mappings:

\begin{align} \label{eq:logical_many}
    &([R_x(\theta)] , [(v_3)] , [\emptyset])
    \\& \qquad\rightarrow ([R_x(\theta), X, X] , [(v_3), (v_1), (v_1)] , [\emptyset, \emptyset, \emptyset]) \nonumber\\
    &([R_x(\theta), X, X] , [(v_3), (v_1), (v_1)] , [\emptyset, \emptyset, \emptyset])\label{eq:hardware_many} 
    \\&\qquad\rightarrow ([R_x(\theta)] , ([v_3]) , [\emptyset]) \nonumber
\end{align}

\noindent Oftentimes logical circuits using elements of $V_L$ with higher degree than those available in $V_p$, sometimes resulting in the need for SWAP networks to realize an operation $U(\mathcal{C}_L)$ on hardware, for example:
\begin{align}\label{eq:swapnetwork_ex}
     &([\text{CNOT}] , [\emptyset] , [(v_1, v_3)])
     \\&\qquad \rightarrow 
     \begin{array}{ccc}
          ([\text{SWAP}, &  \text{CNOT}, &  \text{SWAP}], \\ \left[ \emptyset\right. , &  \emptyset, & \left. \emptyset \right], \\ \left[(v_1, v_2), \right. & (v_2, v_3), & \left.(v_1, v_2)\right])
     \end{array}
\end{align}

\noindent Allowing for mappings of this form complicates the implementation of SA. Compiling operations from $\mathcal{G}_L$ to $\mathcal{G}_p$ or simplifying the implementation of $\mathcal{C}_L$ (e.g. Eq.~\ref{eq:hardware_many}) to reduce the depth or number of operations in a hardware circuit involves a compilation problem that is generally NP-hard \cite{botea2018complexity}.  On the other hand, allowing for arbitrary maps of the form shown in Eqs.~\ref{eq:logical_many} and \ref{eq:swapnetwork_ex} will result in a state space with infinite configurations. In practice this means that the set $S^*$ of optimal qubit assignments may not be accessible by SA or any optimization over qubit assignments. 

\subsection{Assignment of circuits with unconstrained logical connectivity}\label{app:annealer_algorithms}

Algorithm~\ref{alg:sa_finite} describes the implementation of SA for qubit assignment given a neighborhood function $S: \mathbf{C}_p\rightarrow  \mathbf{C}_p$ and a fixed temperature schedule $T: \mathbb{Z}\rightarrow \mathbb{R}$. 
\begin{algorithm}[ht]
\SetAlgoLined
\KwResult{ $\mathcal{C}_p \in S$ }
 initialize $T \leftarrow T(0)$,\, $\mathcal{C}_p$ \;
 compute cost $C =  C_{LE}(\mathcal{C}_p)$ \;
 \For{$i=1\dots N$}{
  sample $\mathcal{C}_p' \in S(\mathcal{C}_p)$ \;
  $C' = C_{LE}(\mathcal{C}_p')$ \;
  \eIf{$C' < C$}{
    update $\mathcal{C}_p \leftarrow \mathcal{C}_p'$, $C \leftarrow C'$ \;
    }{
    sample $x \sim \text{Uniform}(0,1)$ \;
    \If{$x < \exp\left( (C' - C ) /T\right)$
    }{
    update $\mathcal{C}_p \leftarrow \mathcal{C}_p'$, $C \leftarrow C'$ \;
    }
    }
    update $T \leftarrow T(i)$ \;
 }
 \caption{Simulated annealing for qubit assignment, $N$ iterations.}
 \label{alg:sa_finite}
\end{algorithm}

\begin{figure*}
    \centering
    \includegraphics[width=.8\textwidth]{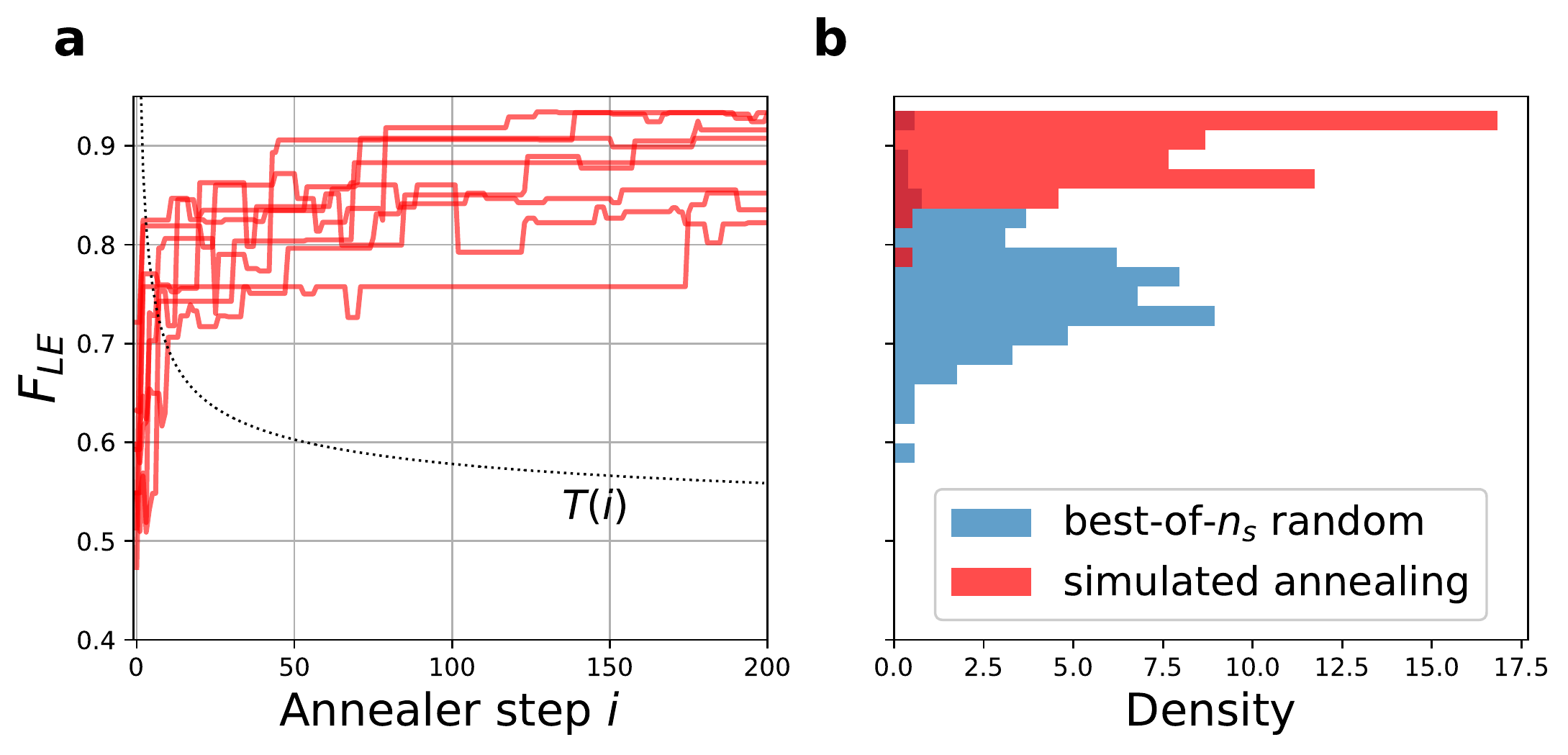}
    \caption{\textbf{a} Sample qubit assignment score at each step of SA for a random 4-qubit circuit with unconstrained logical connectivity. Each $\loschmidtf$ is computed for a qubit assignment initially placed onto four simulated qubits before being transpiled \cite{qiskit} into a circuit configuration satisfying nearest-neighbor connectivity on a $5 \times 5$ 2D grid. A maximum score $\loschmidtf^*$ was not identified due to the number of assignment in the state space (the $\binom{25}{4}4!$ possible all-to-all connectivity initial assignments on the grid result in an unknown number of possible transpiled circuits satisfying nearest neighbor connectivity). Experiments were performed with a tuned logarithmic cooling schedule (arbitrarily rescaled for presentation), $T(i) = T_0 / (1 + \log (1 + i))$ with $T_0 = 0.08$. \textbf{b} 100 trials of SA with transpiled all-to-all connectivity significantly outperforms best-of-$n_s$ random assignment where $n_s$ represents the number of unique states queried by each annealer.}
    \label{fig:sa_alltoall_results}
\end{figure*}

In our simulated experiments involving the assignment of logical circuits with all-to-all logical connectivity to a hardware graph with restricted connectivity, we first assumed $V_L = V_p$ and used a transpiler \cite{qiskit} to construct an intermediate logical circuit $\mathcal{C}_L$ respecting (simulated) hardware connectivity. Then we defined the neighborhood $S(\mathcal{C}_L)$ around $\mathcal{C}_L$ for unconstrained logical connectivity (Sec.~\ref{sec:constrained_1})  according to Algorithm~\ref{alg:sa_alltoall}. Despite not having access to the complete state space of qubit assignments, this update scheme was able to reliably find high scoring qubit assignments in simulation.

\begin{algorithm}[ht]
\SetAlgoLined
\KwResult{ Sample $\mathcal{C}_L' \in S(\mathcal{C}_L)$ }
$\mathcal{C}_L' \leftarrow \mathcal{C}_L$
construct an ordering of qubits $\mathbf{v}$ in $\mathcal{C}_L$\;
sample $x \in \{1, 2, 3\}$ \;
\uIf{$x = 1$
}{
Swap two elements $v_i, v_j \in \mathbf{v}$ \;
}
\uElseIf{$x = 2$
}{
``Shift'' all $v_i \in \mathbf{v}$ by one row or column
}
\uElseIf{$x = 3$
}{
Swap an element $v_i\in \mathbf{v}$ with a qubit in $V_p - \{v_k\}_{k=1}^n$\;
}
Transpile $\mathcal{C}_L'$ onto hardware topology\;
 \caption{An implicit definition for the neighborhood around $\mathcal{C}_L$ in terms of an algorithm applied at each iteration of SA.}
\label{alg:sa_alltoall}
\end{algorithm}

Fig.~\ref{fig:sa_alltoall_results} demonstrates results of running SA for all-to-all connectivity logical circuits transpiled onto a nearest-neighbor connectivity simulated hardware device. While this approach was successful over random assignment of similar circuits, we emphasize that transpilation by default represents a one-to-many mapping and so further work is necessary to constrain the map $M$ induced by transpilation such that $S^*$ remains accessible by SA. One potential approach is to add a counting argument to $M$ that enumerates over all choices of SWAP networks such that $U(\mathcal{C}_p) = U$, and to incorporate this counter into the SA update step. More generally, the task of noise-aware quantum compilation involving optimization over qubit assignments on noisy hardware with a state space $\mathbf{C}_p$ induced by arbitrary $M$ remains an open problem that we leave to future work.

\section{Device specifications and additional experiments}\label{app:circuits}

\begin{figure}
\raggedright
\includegraphics[width=.8\columnwidth]{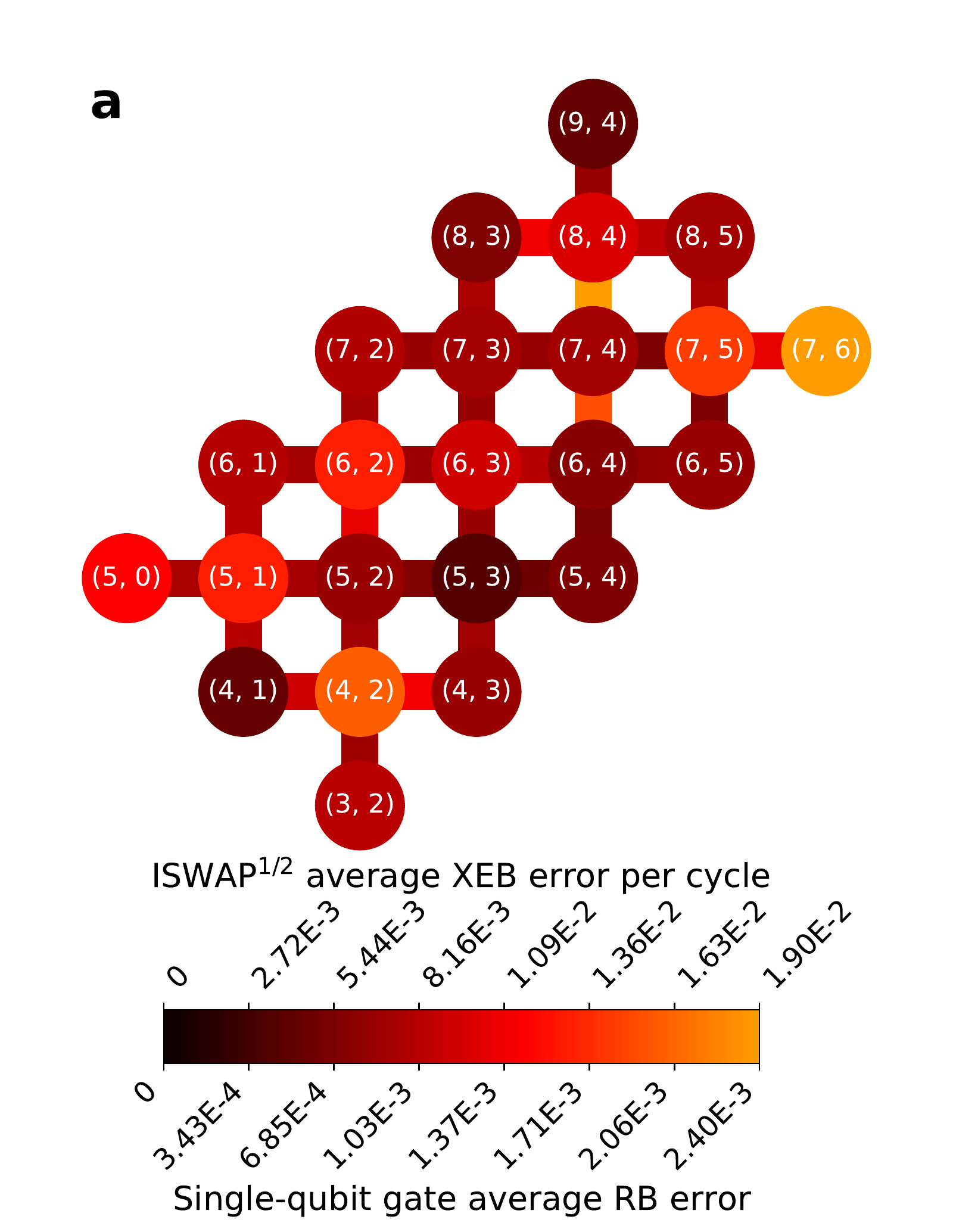} %
\includegraphics[width=\columnwidth]{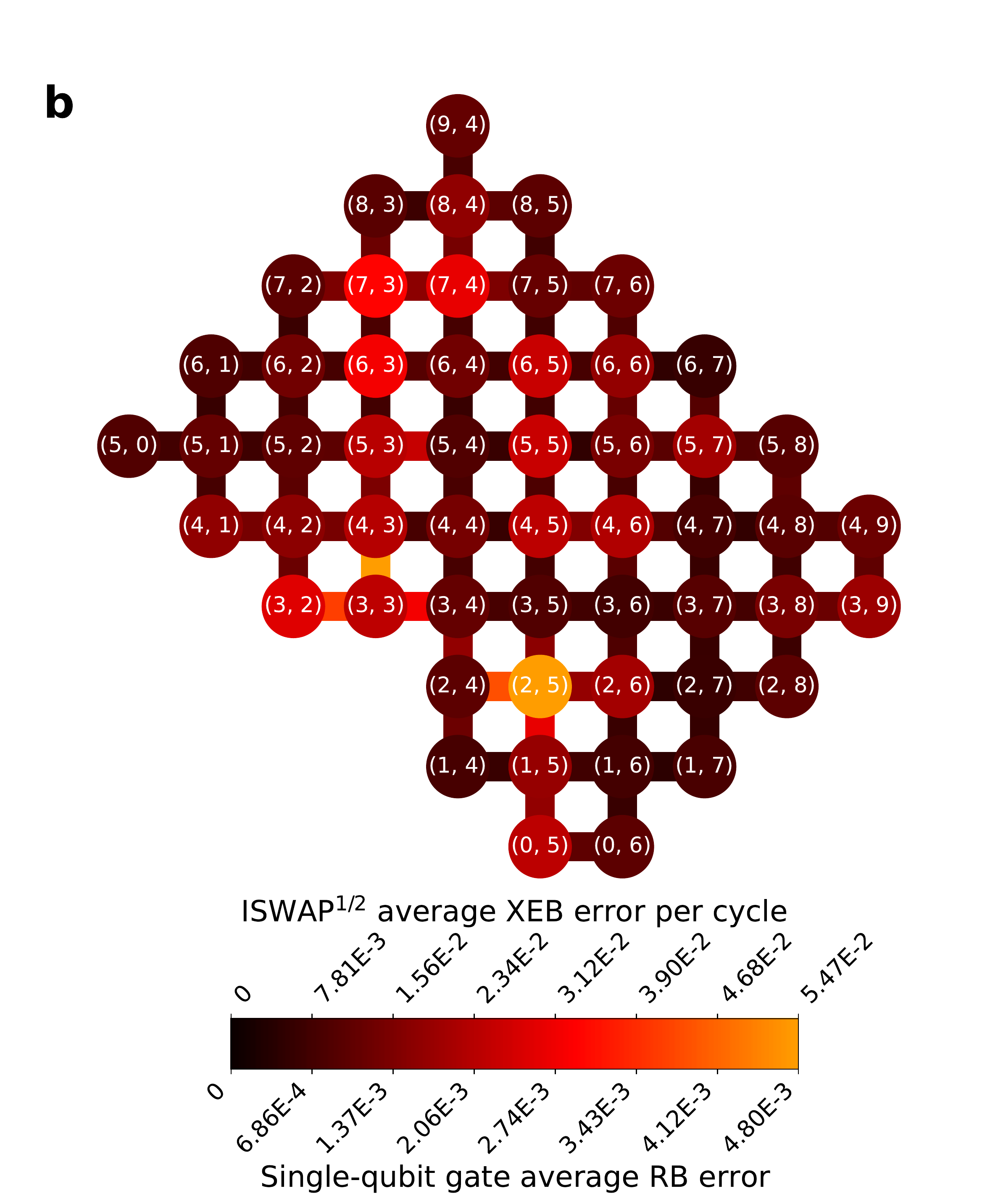}

\caption{Hardware layouts for the \textbf{a} ``Rainbow'' 23-qubit device and \textbf{b} ``Weber'' 53-qubit device accessed via Google's Quantum Computing Service. The color of each node shows single-qubit RB error for gates executed on the corresponding qubit, while the color of each edge shows XEB error per cycle for two-qubit gates executed on the corresponding edge, with darker colors corresponding to lower error rates. Typical error rates were $0.1\%$ and $1\%$ for RB and XEB respectively, and these values were used to construct $\hat{\boldsymbol{\epsilon}}$ and $\hat{\boldsymbol{\eta}}$ to compute $F_0$ in hardware experiments.}
\label{fig:hardware_diagrams}
\end{figure}

We provide detail on the circuits implemented and the hardware devices used for hardware execution. We accessed the Weber and Rainbow superconducting qubit devices through Google's Quantum Computing Service. Fig.~\ref{fig:hardware_diagrams} shows the layout and connectivity of the qubits on each of these devices, colored according to gate fidelity metrics queried from the processor. 

\begin{figure}
    \centering
    \includegraphics[width=\columnwidth]{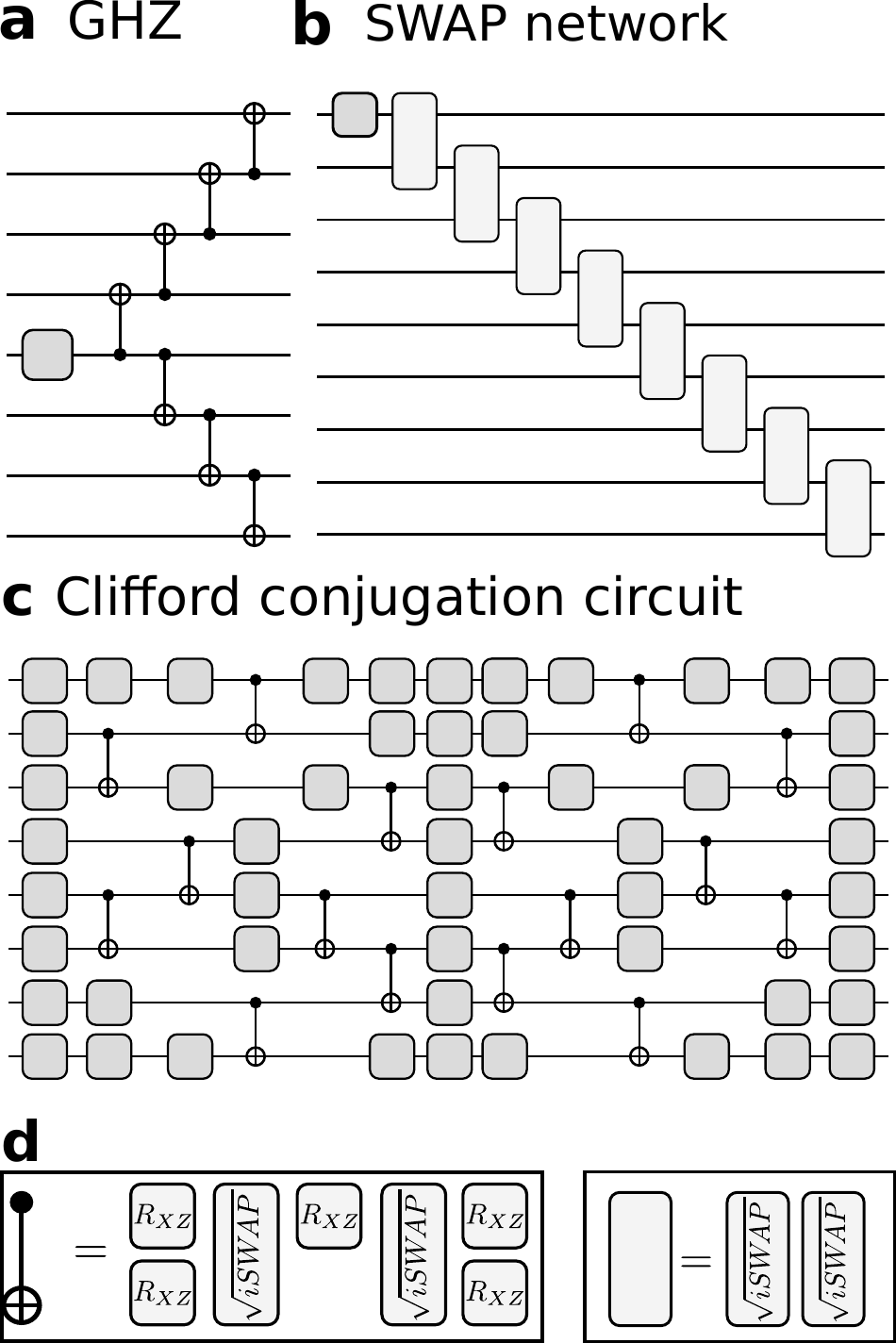}
    \caption{Circuits implemented for hardware diagnostic experiments for $\loschmidtf$. \textbf{a.} The GHZ state $|\psi_{GHZ}\rangle = |0^n\rangle + |1^n\rangle$ can be prepared on a linear connectivity in $O(n)$ depth.  \textbf{b.} The SWAP network circuit prepares a state $|\psi\rangle = \alpha |0\rangle +\beta |1\rangle$ that is then swapped into a new register using $4m$ $\text{iSWAP}$ gates for some integer $m$. The fidelity of this circuit can be computed in constant time by performing the projective measurement with elements $\{ \mathbb{I} - |\psi\rangle\langle \psi|, |\psi\rangle\langle \psi| \}$. \textbf{c.} A sample Clifford conjugation circuit: An arbitrary, random Clifford circuit will typically require significant resources to certify. However the output for circuits of the form $U P U^\dagger$ with $U \in \text{Cl}(2^n)$ and $P \in \{I, X, Y, Z\}^n$ can be verified in constant time. \textbf{d.} Decompositions of entangling gates using hardware native gateset for both Rainbow and Weber devices. $\text{CNOT}$ gates are decomposed into $\sqrt{iSWAP}$ and the $\text{PhasedXZ}(a,b,c)$ gate defined as $R{XZ} = R_z(a) R_x(b) Rz(c)$ and implemented with depth-1 on hardware.}
    \label{fig:my_label}
\end{figure}

\begin{figure*}[t]
    \centering
    \includegraphics[width=\textwidth]{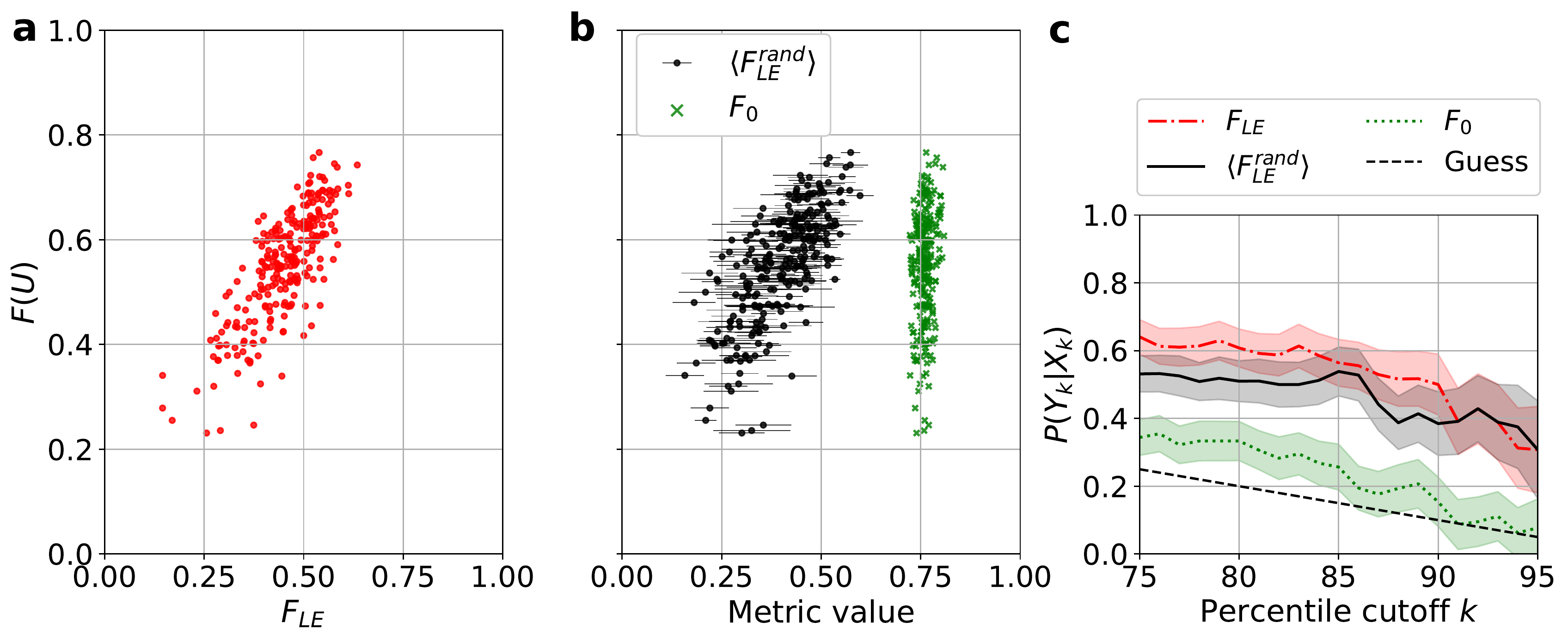}
     \caption{In the case of random Clifford conjugation circuits, computing $\loschmidtf$ for the specific instance of random circuit does not provide significant advantage over  computing $\langle F_{LE}^{rand} \rangle$ using a different kind of random circuit shown in \textbf{b}. Notably, both metrics still greatly outperform $F_0$ which was computed using RB and XEB diagnostic error rates. $F_0$ computed using RB error rates ostensibly describes the fidelity of a random circuit since the effect of Clifford twirling is understood to result in local depolarizing errors, though XEB fidelities do not have a similar straightforward interpretation. \textbf{c} $P(F {>k}|\loschmidtf >k)$ and $P(F {>k}|\langle \loschmidtf^{rand} \rangle {>k})$ do not differ significantly, demonstrating that either metric is equally useful for picking a qubit assignment to optimize $F$. } 
    \label{fig:cliffconj_results}
\end{figure*}

\begin{figure*}[t]
    \centering
    \includegraphics[width=\textwidth]{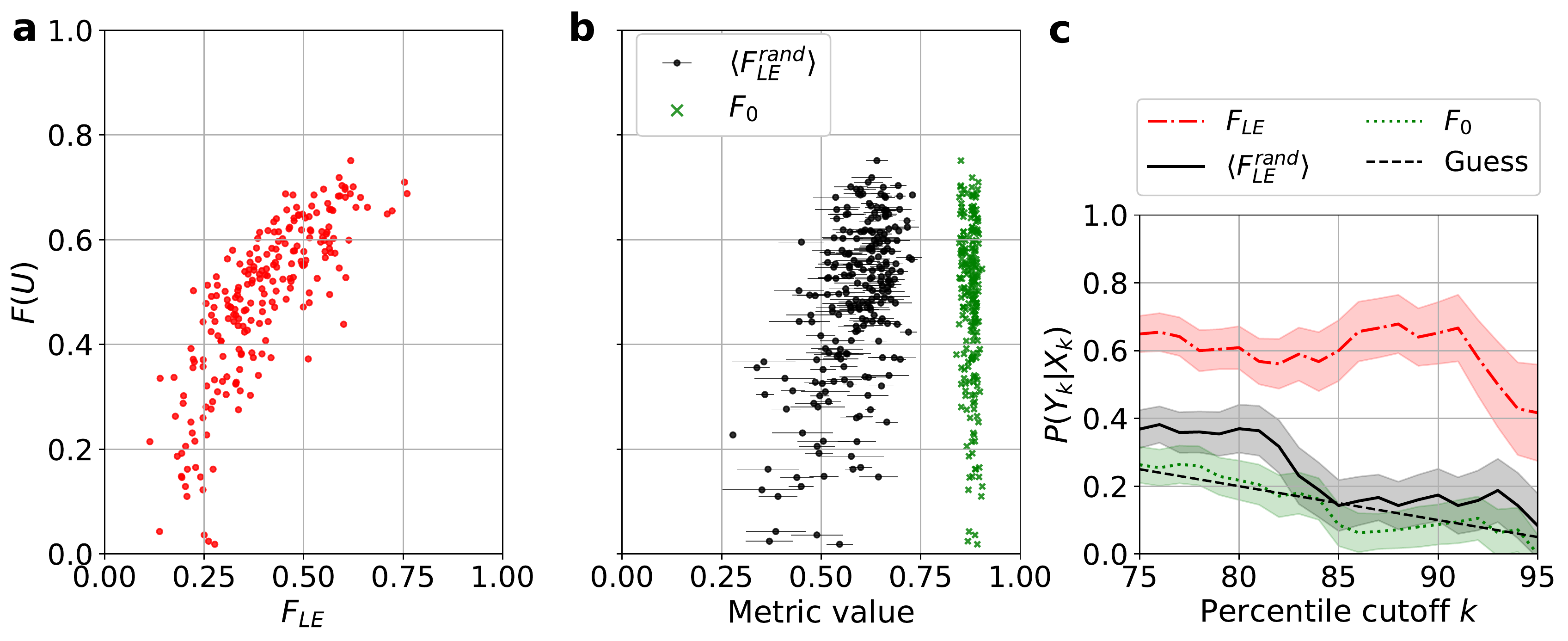}
      \caption{ \textbf{a.} Comparison of $F$ to $F_{LE}$ for a 9-qubit SWAP network experiment. Similarly to the GHZ experiment, the fidelity of the state transmitted by the SWAP network was strongly correlated with $\loschmidtf$ computed on the same qubit assignment. \textbf{b} $F_{LE}^{rand}$ computed using random circuits and $F_0$ computed using calibration data were not strongly correlated with actual device performance.  \textbf{c} The conditional success probability $P(Y_{k}|X_{k})$ further emphasizes the performance of $\loschmidtf$ compared to $F_0$ and $\langle \loschmidtf^{rand} \rangle$.}
      \label{fig:swapnet_results}
\end{figure*}

Each estimator accumulates error due to finite sample effects, gate infidelity in the DFE measurement subcircuit, and residual readout error after readout error correction has been applied. Without detailed knowledge of these effects we cannot completely characterize the error in each diagnostic, and so we assume every experimental diagnostic is affected equally by these systematic errors. Figs.~\ref{fig:cliffconj_results}-\ref{fig:swapnet_results} show the results of diagnostic runs for the (random) Clifford conjugation circuits and SWAPnet circuits respectively. For a specific family of random Clifford circuits we find that  $\loschmidtf$ and $\langle \loschmidtf^{rand}\rangle$ capture the behavior of the fidelity $F$ equally well when the target circuit is itself a random circuit. In contrast, the fidelity of a state transported through a SWAP network is only poorly modeled by $\loschmidtf$ computed on random circuits. In both cases, attempting to find a qubit assignment that maximizes $F$ using knowledge of $F_0$ alone is generally no better than random guessing, and in some cases this strategy is worse than random guessing (Fig.~\ref{fig:swapnet_results}c). 
\begin{figure}[t]
    \centering
     \includegraphics[width=\columnwidth]{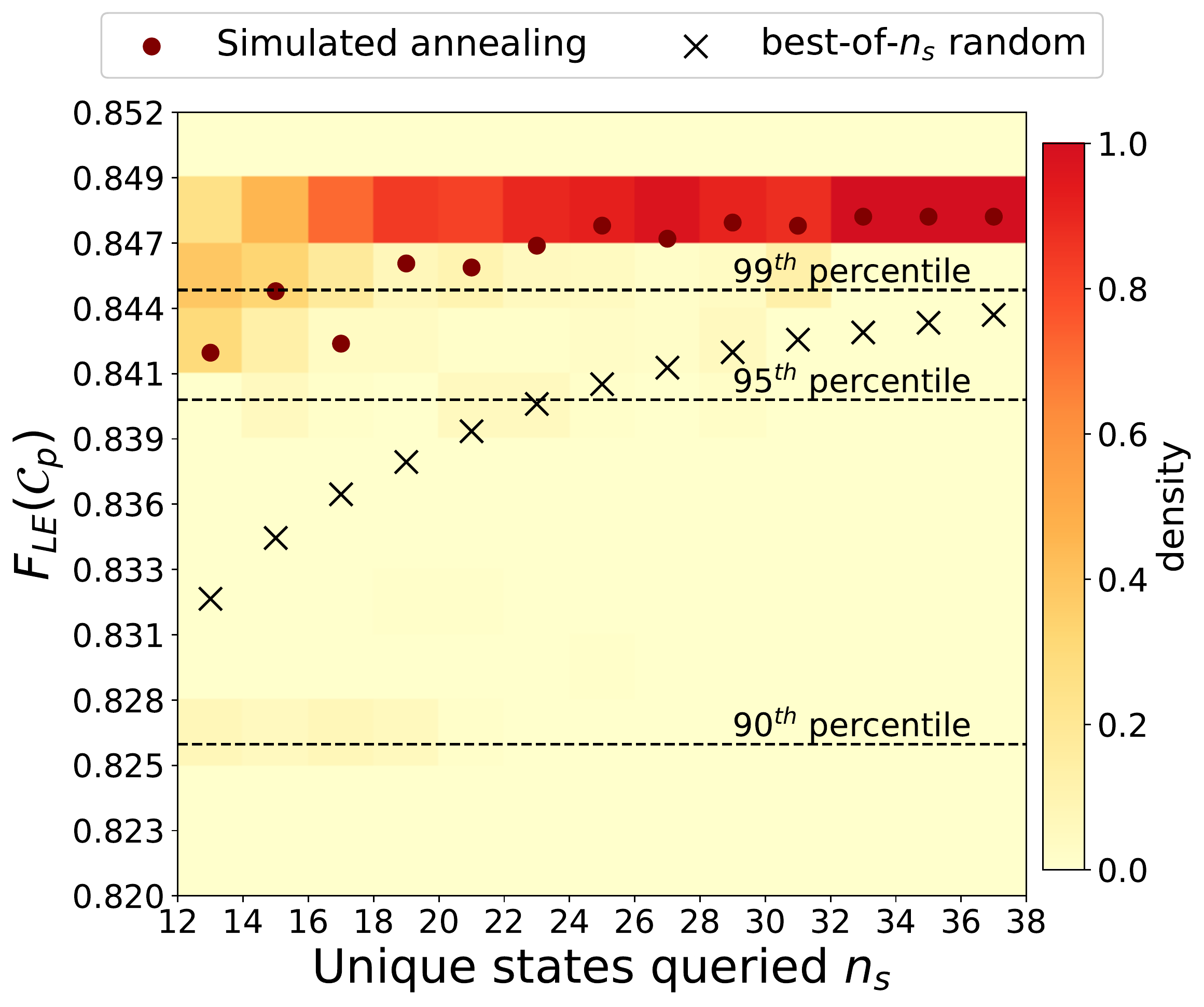}
      \caption{2D histogram of best $\loschmidtf$ values found over 1000 trials of Simulated Annealing performed on $\loschmidtf$ values computed for $U_{QFT}|j\rangle$ with $n=3$ for the complete state space of 148 circuit assignments on Rainbow. Each $\loschmidtf$ is the average over all choices of $j \in [0, 3, 4, 7]$. We used an exponential temperature decay $T_i = T_0 \alpha^i$ ($\alpha = 0.988$, $T_0=0.05$). Averaged over all choices of $n_s$ the final $F_{LE}$ achieved by SA outperformed the random sampling scheme by 1.4\%.}
      \label{fig:hardware_SA_exp_qft}
\end{figure}

\subsection{Quantum Fourier Transform}\label{sec:qft}

In addition to the circuits described in the main text, we performed SA for qubit assignment for an $n$ qubit Quantum Fourier Transform (QFT) circuit $U_{QFT}$ using $n=3$ nearest-neighbor connectivity qubits. In general, the state resulting from applying $U_{QFT}$ an arbitrary initial superposition state will have support on an exponential number of Pauli strings when represented in the Pauli basis, so that DFE can not be performed with better than exponential cost. To work around this, we chose to implement $U_{QFT}$ applied to a specific computational basis state $|j\rangle , \, j \in [0, 2^n - 1]$. This requires only a constant number of experiments since the result of this computation a separable state:
\begin{align}
|\tilde j\rangle = U_{QFT}|j \rangle &= \frac{1}{\sqrt{2^n}}\sum_{k=0}^{2^n - 1} \exp\left(-i2\pi jk 2^{-n}\right)
\\&= \frac{1}{\sqrt{2^n}} \bigotimes_{p=1}^n \left(|0 \rangle +  \exp\left(-i2\pi j 2^{-p}\right)|1\rangle \right)
\end{align}

Defining $|a_p\rangle = 2^{-1/2} \left(|0 \rangle +  \exp\left(-i2\pi j 2^{-p}\right)|1\rangle \right)$ and setting the measurement parameters
\begin{align}
    \theta_p &= 2\pi j 2^{-p} \\
    U_p &= R_z\left(-\theta_p  \right)   \\
    M_p &= U_pH\sigma_z H U_p^\dagger \\
\end{align}

gives the Pauli decomposition for the state over each local system as
\begin{equation}
    |a_p \rangle \langle a_p | = \frac{1}{2} (I + M_p)
\end{equation}

Then given a state $\rho$ output by running $QFT$ on hardware with input $|j\rangle$ we can directly estimate the fidelity of the prepared state as 

\begin{align}
\mathcal{F}(\rho, |\tilde{j}\rangle\langle\tilde{j} |) &= \frac{1}{2^n} \text{Tr}\left( \rho \bigotimes_{p=1}^{n}  \left( I + M_p \right) \right)
\\&= \frac{1}{2^n} \sum_{\ell\in\{0,1\}^n} \langle \mathbf{M}^\ell \rangle 
\end{align}

where $\mathbf{M}^\ell = M_1^{\ell_1} \otimes M_2^{\ell_2} \otimes \dots \otimes M_{n}^{\ell_{n}}$. Therefore a single measurement configuration for the operator $ M_1 \otimes M_2 \otimes \dots \otimes M_{n}$ contains all of the information needed to estimate $F$ up to postprocessing.For each qubit assignment we evaluated $F_{LE}$ and $F$ were evaluated for $j \in [0, 3, 4, 7]$  for the $n=3$ $U_{QFT}|j\rangle$ states on all possible qubit assignments on the Rainbow device. As with other experiments, $\loschmidtf$ with $\tau_b (\loschmidtf, F) = 0.72$ outperforms other metrics considered ($\tau_b (\langle \loschmidtf^{rand}\rangle, F) = 0.45$, $\tau_b(F_0, F) = 0.15$) when used as a proxy for maximizing $F$ over qubit assignments. Fig.~\ref{fig:hardware_SA_exp_qft} shows SA for qubit assignment on the state space of length three simple paths on the 23 qubit Rainbow device (148 possible assignments). The performance of SA over a comparable random sampling scheme was less significant compared to the $n=4$ GHZ qubit assignment. This is likely due to the small size of the state space, for which random guessing is much more likely to produce a performant assignment.


\edit{
\subsection{GHZ State Direct Fidelity Estimation}\label{sec:ghz_dfe}
Here we provide measurement setting specifications for performing DFE on the GHZ state according to \cite{G_hne_2007}. The state $|\psi\rangle = \frac{1}{\sqrt{2}}(|0\rangle^{\otimes n} + |1\rangle^{\otimes n})$ can be expressed as a linear combination of Pauli matrices using the equalities
\begin{align}
|0\rangle \langle 0| &= \frac{1}{2} (1 + \sigma_z) \\
|1\rangle \langle 1| &= \frac{1}{2} (1 - \sigma_z) \\
|0\rangle \langle 1| &= \frac{1}{2} (\sigma_x + i\sigma_y) \\
|1\rangle \langle 0| &= \frac{1}{2} (\sigma_x - i\sigma_y).
\end{align}
The goal is to measure $F = \langle \psi|\rho |\psi\rangle$ for some experimentally prepared $\rho$. This fidelity may be expressed as the sum of two terms, $F \equiv f_{XY} + f_Z$, where we define
\begin{align}
f_Z &=\frac{1}{2}\text{Tr}\left( \rho (|0\rangle \langle 0|^{\otimes n} + |1\rangle \langle 1|^{\otimes n}) \right) \\
&= \frac{1}{2}\text{Tr}\left( \rho \frac{1}{2^n} (1 + \sigma_z)^{\otimes n}\right) + \frac{1}{2}\text{Tr}\left( \rho \frac{1}{2^n} (1 - \sigma_z)^{\otimes n}\right) \\
&= \frac{1}{2^{(n+1)}} \left( \sum_{\ell \in \{0,1\}^n} \langle Z^\ell \rangle (1 + (-1)^{|\ell|} ) \right).
\end{align}
Here, $Z^\ell = \sigma_z^{\ell_1} \otimes \sigma_z^{\ell_2} \otimes \dots \otimes \sigma_z^{\ell_n}$ with $\ell \in \{0,1\}^n$ is a Pauli operator, and the prefactor of $1/2$ is because $\psi$ contains $\frac{1}{2}( |0\rangle \langle 0|^{\otimes n} + |1\rangle \langle 1|^{\otimes n})$. To estimate $f_Z$ it is therefore sufficient to measure in the computational basis for each qubit and perform classical postprocessing. To estimate $f_{XY}$, we begin by defining
\begin{align}
M_k &= \left( \cos\left(\frac{k \pi}{n}\right) \sigma_x + \sin \left(\frac{k \pi}{n}\right) \sigma_y \right)^{\otimes n} \\
&= U_k H \sigma_z H U_k^\dagger
\end{align}
where $k=1, \dots, n$ and $U_k := R_z\left(\frac{k \pi}{n} \right)$. Then one can show that
\begin{align}
f_{XY} &= \frac{1}{2}\text{Tr}\left(\rho (|0\rangle \langle 1|^{\otimes n} + |1\rangle \langle 0|^{\otimes n})\right) \\
&= \frac{1}{2n} \sum_{k=1}^n (-1)^k\text{Tr}(M_k \rho).
\end{align}
Therefore to estimate $f_{XY}$ it is sufficient to estimate $\langle M_k \rangle$ using pre-measurement rotations $H U_k^\dagger$ for each choice of $k=1, \dots, n$. As a consequence we are able to estimate fidelity of the GHZ state $|\psi\rangle$ using $n + 1$ experiments.
}


\clearpage

\begin{thebibliography}{69}%
\makeatletter
\providecommand \@ifxundefined [1]{%
 \@ifx{#1\undefined}
}%
\providecommand \@ifnum [1]{%
 \ifnum #1\expandafter \@firstoftwo
 \else \expandafter \@secondoftwo
 \fi
}%
\providecommand \@ifx [1]{%
 \ifx #1\expandafter \@firstoftwo
 \else \expandafter \@secondoftwo
 \fi
}%
\providecommand \natexlab [1]{#1}%
\providecommand \enquote  [1]{``#1''}%
\providecommand \bibnamefont  [1]{#1}%
\providecommand \bibfnamefont [1]{#1}%
\providecommand \citenamefont [1]{#1}%
\providecommand \href@noop [0]{\@secondoftwo}%
\providecommand \href [0]{\begingroup \@sanitize@url \@href}%
\providecommand \@href[1]{\@@startlink{#1}\@@href}%
\providecommand \@@href[1]{\endgroup#1\@@endlink}%
\providecommand \@sanitize@url [0]{\catcode `\\12\catcode `\$12\catcode
  `\&12\catcode `\#12\catcode `\^12\catcode `\_12\catcode `\%12\relax}%
\providecommand \@@startlink[1]{}%
\providecommand \@@endlink[0]{}%
\providecommand \url  [0]{\begingroup\@sanitize@url \@url }%
\providecommand \@url [1]{\endgroup\@href {#1}{\urlprefix }}%
\providecommand \urlprefix  [0]{URL }%
\providecommand \Eprint [0]{\href }%
\providecommand \doibase [0]{https://doi.org/}%
\providecommand \selectlanguage [0]{\@gobble}%
\providecommand \bibinfo  [0]{\@secondoftwo}%
\providecommand \bibfield  [0]{\@secondoftwo}%
\providecommand \translation [1]{[#1]}%
\providecommand \BibitemOpen [0]{}%
\providecommand \bibitemStop [0]{}%
\providecommand \bibitemNoStop [0]{.\EOS\space}%
\providecommand \EOS [0]{\spacefactor3000\relax}%
\providecommand \BibitemShut  [1]{\csname bibitem#1\endcsname}%
\let\auto@bib@innerbib\@empty

\bibitem{Preskill_2018}
John Preskill.
\newblock Quantum computing in the nisq era and beyond.
\newblock {\em Quantum}, 2:79, Aug 2018.

\bibitem{finigan_qubit_2018}
Will Finigan, Michael Cubeddu, Thomas Lively, Johannes Flick, and Prineha
  Narang.
\newblock Qubit allocation for noisy intermediate-scale quantum computers.
\newblock {\em arXiv:1810.08291}, October 2018.

\bibitem{tannu_not_2019}
Swamit~S. Tannu and Moinuddin~K. Qureshi.
\newblock Not all qubits are created equal.
\newblock In {\em Proceedings of the Twenty-Fourth International Conference on
  Architectural Support for Programming Languages and Operating Systems}, pages
  987--999, Providence RI USA, April 2019. ACM.

\bibitem{Note1}
In the context of this work, ``logical qubits'' are an abstraction used in the
  specification of a quantum circuit but have no correspondence to any physical
  device, while ``physical qubits'' exist on a specific device and are subject
  to connectivity constraints and errors. Qubit assignment is then the
  procedure each logical qubit This differs from the logical-vs-physical
  distinction typically used in the context of quantum error correction (QEC).

\bibitem{Emerson_2005}
Joseph Emerson, Robert Alicki, and Karol Życzkowski.
\newblock Scalable noise estimation with random unitary operators.
\newblock {\em J. opt. B Quantum semiclass. opt.}, 7(10):347--–352, Sep 2005.

\bibitem{Knill_2008}
E.~Knill, D.~Leibfried, R.~Reichle, J.~Britton, R.~B. Blakestad, J.~D. Jost,
  C.~Langer, R.~Ozeri, S.~Seidelin, and D.~J. Wineland.
\newblock Randomized benchmarking of quantum gates.
\newblock {\em Phys. Rev. A}, 77(1), Jan 2008.

\bibitem{Boixo_2018}
Sergio Boixo, Sergei~V. Isakov, Vadim~N. Smelyanskiy, Ryan Babbush, Nan Ding,
  Zhang Jiang, Michael~J. Bremner, John~M. Martinis, and Hartmut Neven.
\newblock Characterizing quantum supremacy in near-term devices.
\newblock {\em Nature Physics}, 14(6):595–600, Apr 2018.

\bibitem{Neill_2018}
C.~Neill, P.~Roushan, K.~Kechedzhi, S.~Boixo, S.~V. Isakov, V.~Smelyanskiy,
  A.~Megrant, B.~Chiaro, A.~Dunsworth, K.~Arya, and et~al.
\newblock A blueprint for demonstrating quantum supremacy with superconducting
  qubits.
\newblock {\em Science}, 360(6385):195–199, Apr 2018.

\bibitem{Chen2021}
Zijun Chen, Kevin~J. Satzinger, Juan Atalaya, Alexander~N. Korotkov, Andrew
  Dunsworth, Daniel Sank, Chris Quintana, Matt McEwen, Rami Barends, Paul~V.
  Klimov, Sabrina Hong, Cody Jones, Andre Petukhov, Dvir Kafri, Sean Demura,
  Brian Burkett, Craig Gidney, Austin~G. Fowler, Alexandru Paler, Harald
  Putterman, Igor Aleiner, Frank Arute, Kunal Arya, Ryan Babbush, Joseph~C.
  Bardin, Andreas Bengtsson, Alexandre Bourassa, Michael Broughton, Bob~B.
  Buckley, David~A. Buell, Nicholas Bushnell, Benjamin Chiaro, Roberto Collins,
  William Courtney, Alan~R. Derk, Daniel Eppens, Catherine Erickson, Edward
  Farhi, Brooks Foxen, Marissa Giustina, Ami Greene, Jonathan~A. Gross,
  Matthew~P. Harrigan, Sean~D. Harrington, Jeremy Hilton, Alan Ho, Trent Huang,
  William~J. Huggins, L.~B. Ioffe, Sergei~V. Isakov, Evan Jeffrey, Zhang Jiang,
  Kostyantyn Kechedzhi, Seon Kim, Alexei Kitaev, Fedor Kostritsa, David
  Landhuis, Pavel Laptev, Erik Lucero, Orion Martin, Jarrod~R. McClean, Trevor
  McCourt, Xiao Mi, Kevin~C. Miao, Masoud Mohseni, Shirin Montazeri, Wojciech
  Mruczkiewicz, Josh Mutus, Ofer Naaman, Matthew Neeley, Charles Neill, Michael
  Newman, Murphy~Yuezhen Niu, Thomas~E. O'Brien, Alex Opremcak, Eric Ostby,
  B{\'a}lint Pat{\'o}, Nicholas Redd, Pedram Roushan, Nicholas~C. Rubin,
  Vladimir Shvarts, Doug Strain, Marco Szalay, Matthew~D. Trevithick, Benjamin
  Villalonga, Theodore White, Z.~Jamie Yao, Ping Yeh, Juhwan Yoo, Adam Zalcman,
  Hartmut Neven, Sergio Boixo, Vadim Smelyanskiy, Yu~Chen, Anthony Megrant,
  Julian Kelly, and Google~Quantum AI.
\newblock Exponential suppression of bit or phase errors with cyclic error
  correction.
\newblock {\em Nature}, 595(7867):383--387, Jul 2021.

\bibitem{McEwen2021}
M.~McEwen, D.~Kafri, Z.~Chen, J.~Atalaya, K.~J. Satzinger, C.~Quintana, P.~V.
  Klimov, D.~Sank, C.~Gidney, A.~G. Fowler, F.~Arute, K.~Arya, B.~Buckley,
  B.~Burkett, N.~Bushnell, B.~Chiaro, R.~Collins, S.~Demura, A.~Dunsworth,
  C.~Erickson, B.~Foxen, M.~Giustina, T.~Huang, S.~Hong, E.~Jeffrey, S.~Kim,
  K.~Kechedzhi, F.~Kostritsa, P.~Laptev, A.~Megrant, X.~Mi, J.~Mutus,
  O.~Naaman, M.~Neeley, C.~Neill, M.~Niu, A.~Paler, N.~Redd, P.~Roushan, T.~C.
  White, J.~Yao, P.~Yeh, A.~Zalcman, Yu~Chen, V.~N. Smelyanskiy, John~M.
  Martinis, H.~Neven, J.~Kelly, A.~N. Korotkov, A.~G. Petukhov, and R.~Barends.
\newblock Removing leakage-induced correlated errors in superconducting quantum
  error correction.
\newblock {\em Nature Communications}, 12(1):1761, Mar 2021.

\bibitem{PhysRevLett.128.110504}
Edward~H. Chen, Theodore~J. Yoder, Youngseok Kim, Neereja Sundaresan, Srikanth
  Srinivasan, Muyuan Li, Antonio~D. C\'orcoles, Andrew~W. Cross, and Maika
  Takita.
\newblock Calibrated decoders for experimental quantum error correction.
\newblock {\em Phys. Rev. Lett.}, 128:110504, Mar 2022.

\bibitem{PhysRevLett.129.030501}
Youwei Zhao, Yangsen Ye, He-Liang Huang, Yiming Zhang, Dachao Wu, Huijie Guan,
  Qingling Zhu, Zuolin Wei, Tan He, Sirui Cao, Fusheng Chen, Tung-Hsun Chung,
  Hui Deng, Daojin Fan, Ming Gong, Cheng Guo, Shaojun Guo, Lianchen Han, Na~Li,
  Shaowei Li, Yuan Li, Futian Liang, Jin Lin, Haoran Qian, Hao Rong, Hong Su,
  Lihua Sun, Shiyu Wang, Yulin Wu, Yu~Xu, Chong Ying, Jiale Yu, Chen Zha, Kaili
  Zhang, Yong-Heng Huo, Chao-Yang Lu, Cheng-Zhi Peng, Xiaobo Zhu, and Jian-Wei
  Pan.
\newblock Realization of an error-correcting surface code with superconducting
  qubits.
\newblock {\em Phys. Rev. Lett.}, 129:030501, Jul 2022.

\bibitem{https://doi.org/10.48550/arxiv.2207.06431}
Rajeev Acharya, Igor Aleiner, Richard Allen, Trond~I. Andersen, Markus Ansmann,
  Frank Arute, Kunal Arya, Abraham Asfaw, Juan Atalaya, Ryan Babbush, Dave
  Bacon, Joseph~C. Bardin, Joao Basso, Andreas Bengtsson, Sergio Boixo, Gina
  Bortoli, Alexandre Bourassa, Jenna Bovaird, Leon Brill, Michael Broughton,
  Bob~B. Buckley, David~A. Buell, Tim Burger, Brian Burkett, Nicholas Bushnell,
  Yu~Chen, Zijun Chen, Ben Chiaro, Josh Cogan, Roberto Collins, Paul Conner,
  William Courtney, Alexander~L. Crook, Ben Curtin, Dripto~M. Debroy, Alexander
  Del~Toro Barba, Sean Demura, Andrew Dunsworth, Daniel Eppens, Catherine
  Erickson, Lara Faoro, Edward Farhi, Reza Fatemi, Leslie~Flores Burgos,
  Ebrahim Forati, Austin~G. Fowler, Brooks Foxen, William Giang, Craig Gidney,
  Dar Gilboa, Marissa Giustina, Alejandro~Grajales Dau, Jonathan~A. Gross,
  Steve Habegger, Michael~C. Hamilton, Matthew~P. Harrigan, Sean~D. Harrington,
  Oscar Higgott, Jeremy Hilton, Markus Hoffmann, Sabrina Hong, Trent Huang,
  Ashley Huff, William~J. Huggins, Lev~B. Ioffe, Sergei~V. Isakov, Justin
  Iveland, Evan Jeffrey, Zhang Jiang, Cody Jones, Pavol Juhas, Dvir Kafri,
  Kostyantyn Kechedzhi, Julian Kelly, Tanuj Khattar, Mostafa Khezri, Mária
  Kieferová, Seon Kim, Alexei Kitaev, Paul~V. Klimov, Andrey~R. Klots,
  Alexander~N. Korotkov, Fedor Kostritsa, John~Mark Kreikebaum, David Landhuis,
  Pavel Laptev, Kim-Ming Lau, Lily Laws, Joonho Lee, Kenny Lee, Brian~J.
  Lester, Alexander Lill, Wayne Liu, Aditya Locharla, Erik Lucero, Fionn~D.
  Malone, Jeffrey Marshall, Orion Martin, Jarrod~R. McClean, Trevor Mccourt,
  Matt McEwen, Anthony Megrant, Bernardo~Meurer Costa, Xiao Mi, Kevin~C. Miao,
  Masoud Mohseni, Shirin Montazeri, Alexis Morvan, Emily Mount, Wojciech
  Mruczkiewicz, Ofer Naaman, Matthew Neeley, Charles Neill, Ani Nersisyan,
  Hartmut Neven, Michael Newman, Jiun~How Ng, Anthony Nguyen, Murray Nguyen,
  Murphy~Yuezhen Niu, Thomas~E. O'Brien, Alex Opremcak, John Platt, Andre
  Petukhov, Rebecca Potter, Leonid~P. Pryadko, Chris Quintana, Pedram Roushan,
  Nicholas~C. Rubin, Negar Saei, Daniel Sank, Kannan Sankaragomathi, Kevin~J.
  Satzinger, Henry~F. Schurkus, Christopher Schuster, Michael~J. Shearn, Aaron
  Shorter, Vladimir Shvarts, Jindra Skruzny, Vadim Smelyanskiy, W.~Clarke
  Smith, George Sterling, Doug Strain, Marco Szalay, Alfredo Torres, Guifre
  Vidal, Benjamin Villalonga, Catherine~Vollgraff Heidweiller, Theodore White,
  Cheng Xing, Z.~Jamie Yao, Ping Yeh, Juhwan Yoo, Grayson Young, Adam Zalcman,
  Yaxing Zhang, and Ningfeng Zhu.
\newblock Suppressing quantum errors by scaling a surface code logical qubit,
  2022.

\bibitem{PhysRevA.30.1610}
Asher Peres.
\newblock Stability of quantum motion in chaotic and regular systems.
\newblock {\em Phys. Rev. A}, 30:1610--1615, Oct 1984.

\bibitem{van1987simulated}
Peter~JM Van~Laarhoven and Emile~HL Aarts.
\newblock Simulated annealing.
\newblock In {\em Simulated annealing: Theory and applications}. Springer,
  1987.

\bibitem{proctor2020measuring}
Timothy Proctor, Kenneth Rudinger, Kevin Young, Erik Nielsen, and Robin
  Blume-Kohout.
\newblock Measuring the capabilities of quantum computers.
\newblock {\em Nat. Phys.}, 2021.

\bibitem{proctor2021scalable}
Timothy Proctor, Stefan Seritan, Kenneth Rudinger, Erik Nielsen, Robin
  Blume-Kohout, and Kevin Young.
\newblock Scalable randomized benchmarking of quantum computers using mirror
  circuits, 2021.

\bibitem{gottesman1998heisenberg}
Daniel Gottesman.
\newblock The {Heisenberg} representation of quantum computers.
\newblock In {\em Proceedings of the XXII International Colloquium on Group
  Theoretical Methods in Physics}, pages 32--43, 1998.

\bibitem{gottesman_aaronson_2004}
Scott Aaronson and Daniel Gottesman.
\newblock Improved simulation of stabilizer circuits.
\newblock {\em Phys. Rev. A}, 70(5), Nov 2004.

\bibitem{Nishio_2020}
Shin Nishio, Yulu Pan, Takahiko Satoh, Hideharu Amano, and Rodney~Van Meter.
\newblock Extracting success from {IBM}’s 20-qubit machines using error-aware
  compilation.
\newblock {\em ACM J. Emerg. Technol. Comput.}, 16(3):1–25, Jul 2020.

\bibitem{maslov2008}
Dmitri Maslov, Sean~M. Falconer, and Michele Mosca.
\newblock Quantum circuit placement.
\newblock {\em IEEE Trans. Comput.-Aided Des. Integr. Circuits Syst.},
  27(4):752–763, Apr 2008.

\bibitem{childs2019circuit}
Andrew~M. Childs, Eddie Schoute, and Cem~M. Unsal.
\newblock {Circuit Transformations for Quantum Architectures}.
\newblock In {\em 14th Conference on the Theory of Quantum Computation,
  Communication and Cryptography}, volume 135, pages 3:1--3:24, Dagstuhl,
  Germany, 2019. Schloss Dagstuhl--Leibniz-Zentrum fuer Informatik.

\bibitem{10.1145/3316781.3317859}
Robert Wille, Lukas Burgholzer, and Alwin Zulehner.
\newblock Mapping quantum circuits to ibm qx architectures using the minimal
  number of swap and h operations.
\newblock In {\em Proceedings of the 56th Annual Design Automation Conference
  2019}, DAC '19, New York, NY, USA, 2019. Association for Computing Machinery.

\bibitem{tan2020optimal}
Bochen Tan and Jason Cong.
\newblock Optimal layout synthesis for quantum computing.
\newblock In {\em 2020 IEEE/ACM International Conference On Computer Aided
  Design (ICCAD)}, pages 1--9. IEEE, 2020.

\bibitem{gerard2021string}
Blake Gerard and Martin Kong.
\newblock String abstractions for qubit mapping, 2021.

\bibitem{Zhou_2020}
Xiangzhen Zhou, Sanjiang Li, and Yuan Feng.
\newblock Quantum circuit transformation based on simulated annealing and
  heuristic search.
\newblock {\em IEEE Trans. Comput.-Aided Des. Integr. Circuits Syst.},
  39(12):4683–4694, Dec 2020.

\bibitem{10.1145/3168822}
Marcos~Yukio Siraichi, Vin\'{\i}cius Fernandes~dos Santos, Sylvain Collange,
  and Fernando Magno~Quintao Pereira.
\newblock Qubit allocation.
\newblock In {\em Proceedings of the 2018 International Symposium on Code
  Generation and Optimization}, CGO 2018, page 113–125, New York, NY, USA,
  2018. Association for Computing Machinery.

\bibitem{kitaev1997}
A~Yu Kitaev.
\newblock Quantum computations: algorithms and error correction.
\newblock {\em Russ. Math. Surv.}, 52(6):1191--1249, dec 1997.

\bibitem{qiskit}
H{\'e}ctor Abraham et~al.
\newblock Qiskit: An open-source framework for quantum computing, 2019.

\bibitem{cirq}
{Cirq Developers}.
\newblock Cirq, May 2021.
\newblock {See full list of authors on Github: https://github
  .com/quantumlib/Cirq/graphs/contributors}.

\bibitem{nielsen2002quantum}
Michael~A Nielsen and Isaac Chuang.
\newblock Quantum computation and quantum information, 2002.

\bibitem{PhysRevLett.107.210404}
Marcus~P. da~Silva, Olivier Landon-Cardinal, and David Poulin.
\newblock Practical characterization of quantum devices without tomography.
\newblock {\em Phys. Rev. Lett.}, 107:210404, Nov 2011.

\bibitem{PhysRevLett.106.230501}
Steven~T. Flammia and Yi-Kai Liu.
\newblock Direct fidelity estimation from few pauli measurements.
\newblock {\em Phys. Rev. Lett.}, 106:230501, Jun 2011.

\bibitem{huang2020predicting}
Hsin-Yuan Huang, Richard Kueng, and John Preskill.
\newblock Predicting many properties of a quantum system from very few
  measurements.
\newblock {\em Nat, Phys.}, 16(10):1050--1057, 2020.

\bibitem{loschmidt1876}
J.~Loschmidt.
\newblock Über den zustand des wärmegleichgewichts eines systems von körpern
  mit rücksicht auf die schwerkraft.
\newblock {\em Sitzungsberichte der Akademie der Wissenschaften, Wien},
  II(73):128--142, 1876.

\bibitem{loschmidt_review}
Arseni Wisniacki.
\newblock Loschmidt echo, 2012.

\bibitem{Neeley_2010}
Matthew Neeley, Radoslaw~C. Bialczak, M.~Lenander, E.~Lucero, Matteo
  Mariantoni, A.~D. O’Connell, D.~Sank, H.~Wang, M.~Weides, J.~Wenner, and
  et~al.
\newblock Generation of three-qubit entangled states using superconducting
  phase qubits.
\newblock {\em Nature}, 467(7315):570–573, Sep 2010.

\bibitem{Dewes_2012}
A.~Dewes, F.~R. Ong, V.~Schmitt, R.~Lauro, N.~Boulant, P.~Bertet, D.~Vion, and
  D.~Esteve.
\newblock Characterization of a two-transmon processor with individual
  single-shot qubit readout.
\newblock {\em Phys. Rev. Lett.}, 108(5), Feb 2012.

\bibitem{nachman_unfolding_2020}
Benjamin Nachman, Miroslav Urbanek, Wibe~A de~Jong, and Christian~W Bauer.
\newblock Unfolding quantum computer readout noise.
\newblock {\em npj Quantum Inf.}, 6(1):1--7, 2020.

\bibitem{bravyi2020mitigating}
Sergey Bravyi, Sarah Sheldon, Abhinav Kandala, David~C Mckay, and Jay~M
  Gambetta.
\newblock Mitigating measurement errors in multiqubit experiments.
\newblock {\em Phys. Rev. A}, 103(4):042605, 2021.

\bibitem{PRXQuantum.2.040326}
Paul~D. Nation, Hwajung Kang, Neereja Sundaresan, and Jay~M. Gambetta.
\newblock Scalable mitigation of measurement errors on quantum computers.
\newblock {\em PRX Quantum}, 2:040326, Nov 2021.

\bibitem{peters2021perturbative}
Evan Peters, Andy C.~Y. Li, and Gabriel~N. Perdue.
\newblock Perturbative readout error mitigation for near term quantum
  computers, 2021.

\bibitem{arute2019quantum}
Frank Arute et~al.
\newblock Quantum supremacy using a programmable superconducting processor.
\newblock {\em Nature}, 574(7779):505--510, 2019.

\bibitem{murali2019noise}
Prakash Murali, Jonathan~M Baker, Ali Javadi-Abhari, Frederic~T Chong, and
  Margaret Martonosi.
\newblock Noise-adaptive compiler mappings for noisy intermediate-scale quantum
  computers.
\newblock In {\em Proceedings of the Twenty-Fourth International Conference on
  Architectural Support for Programming Languages and Operating Systems}, pages
  1015--1029, 2019.

\bibitem{peters2021machine}
Evan Peters, João Caldeira, Alan Ho, Stefan Leichenauer, Masoud Mohseni,
  Hartmut Neven, Panagiotis Spentzouris, Doug Strain, and Gabriel~N. Perdue.
\newblock Machine learning of high dimensional data on a noisy quantum
  processor.
\newblock {\em npj Quantum Inf.}, 7(1):1--7, 2021.
\newblock arXiv:2101.09581.

\bibitem{niu2020}
Siyuan Niu, Adrien Suau, Gabriel Staffelbach, and Aida Todri-Sanial.
\newblock A hardware-aware heuristic for the qubit mapping problem in the nisq
  era.
\newblock {\em IEEE Transactions on Quantum Engineering}, 1:1–14, 2020.

\bibitem{8980312}
Prakash Murali, Norbert~Matthias Linke, Margaret Martonosi, Ali~Javadi Abhari,
  Nhung~Hong Nguyen, and Cinthia~Huerta Alderete.
\newblock Full-stack, real-system quantum computer studies: Architectural
  comparisons and design insights.
\newblock In {\em 2019 ACM/IEEE 46th Annual International Symposium on Computer
  Architecture (ISCA)}, pages 527--540, 2019.

\bibitem{PhysRevX.9.021045}
Kenneth Rudinger, Timothy Proctor, Dylan Langharst, Mohan Sarovar, Kevin Young,
  and Robin Blume-Kohout.
\newblock Probing context-dependent errors in quantum processors.
\newblock {\em Phys. Rev. X}, 9:021045, Jun 2019.

\bibitem{Magesan_2011}
Easwar Magesan, J.~M. Gambetta, and Joseph Emerson.
\newblock Scalable and robust randomized benchmarking of quantum processes.
\newblock {\em Phys. Rev. Lett.}, 106(18), May 2011.

\bibitem{Magesan_2012}
Easwar Magesan, Jay~M. Gambetta, and Joseph Emerson.
\newblock Characterizing quantum gates via randomized benchmarking.
\newblock {\em Phys. Rev. A}, 85(4), Apr 2012.

\bibitem{liu2021benchmarking}
Yunchao Liu, Matthew Otten, Roozbeh Bassirianjahromi, Liang Jiang, and Bill
  Fefferman.
\newblock Benchmarking near-term quantum computers via random circuit sampling,
  2021.

\bibitem{bertsimas1993simulated}
Dimitris Bertsimas, John Tsitsiklis, et~al.
\newblock Simulated annealing.
\newblock {\em Statist. Sci.}, 8(1):10--15, 1993.

\bibitem{hajek1988cooling}
Bruce Hajek.
\newblock Cooling schedules for optimal annealing.
\newblock {\em Math. Oper. Res.}, 13(2):311--329, 1988.

\bibitem{holmes_impact_2020}
Adam Holmes, Sonika Johri, Gian~Giacomo Guerreschi, James~S. Clarke, and A.~Y.
  Matsuura.
\newblock Impact of qubit connectivity on quantum algorithm performance.
\newblock {\em Quantum Sci. Technol.}, 5(2):025009, March 2020.

\bibitem{G_hne_2007}
Otfried Gühne, Chao-Yang Lu, Wei-Bo Gao, and Jian-Wei Pan.
\newblock Toolbox for entanglement detection and fidelity estimation.
\newblock {\em Phys. Rev. A}, 76(3), Sep 2007.

\bibitem{isakov2021simulations}
Sergei~V. Isakov, Dvir Kafri, Orion Martin, Catherine~Vollgraff Heidweiller,
  Wojciech Mruczkiewicz, Matthew~P. Harrigan, Nicholas~C. Rubin, Ross Thomson,
  Michael Broughton, Kevin Kissell, Evan Peters, Erik Gustafson, Andy C.~Y. Li,
  Henry Lamm, Gabriel Perdue, Alan~K. Ho, Doug Strain, and Sergio Boixo.
\newblock Simulations of quantum circuits with approximate noise using qsim and
  cirq, 2021.

\bibitem{google_qcs}
Calibration metrics.
\newblock \url{https://quantumai.google/cirq/google/calibration}, 2021.

\bibitem{kendall1938new}
Maurice~G Kendall.
\newblock A new measure of rank correlation.
\newblock {\em Biometrika}, 30(1/2):81--93, 1938.

\bibitem{efron1994introduction}
Bradley Efron and Robert~J Tibshirani.
\newblock {\em An introduction to the bootstrap}.
\newblock CRC press, 1994.

\bibitem{networkx}
Aric~A. Hagberg, Daniel~A. Schult, and Pieter~J. Swart.
\newblock Exploring network structure, dynamics, and function using networkx.
\newblock In {\em Proceedings of the 7th Python in Science Conference}, pages
  11 -- 15, Pasadena, CA USA, 2008.

\bibitem{Harper_2020}
Robin Harper, Steven~T. Flammia, and Joel~J. Wallman.
\newblock Efficient learning of quantum noise.
\newblock {\em Nature Physics}, 16(12):1184--1188, aug 2020.

\bibitem{flammia_2021}
Steven~T. Flammia.
\newblock Averaged circuit eigenvalue sampling, 2021.

\bibitem{Neill2021}
C.~Neill, T.~McCourt, X.~Mi, Z.~Jiang, M.~Y. Niu, W.~Mruczkiewicz, et~al.
\newblock Accurately computing the electronic properties of a quantum ring.
\newblock {\em Nature}, 594(7864):508--512, Jun 2021.

\bibitem{Kandala2019}
Abhinav Kandala, Kristan Temme, Antonio~D. C{\'o}rcoles, Antonio Mezzacapo,
  Jerry~M. Chow, and Jay~M. Gambetta.
\newblock Error mitigation extends the computational reach of a noisy quantum
  processor.
\newblock {\em Nature}, 567(7749):491--495, Mar 2019.

\bibitem{doi:10.1126/sciadv.aaw5686}
Chao Song, Jing Cui, H.~Wang, J.~Hao, H.~Feng, and Ying Li.
\newblock Quantum computation with universal error mitigation on a
  superconducting quantum processor.
\newblock {\em Science Advances}, 5(9):eaaw5686, 2019.

\bibitem{Khatri_2019}
Sumeet Khatri, Ryan LaRose, Alexander Poremba, Lukasz Cincio, Andrew~T.
  Sornborger, and Patrick~J. Coles.
\newblock Quantum-assisted quantum compiling.
\newblock {\em Quantum}, 3:140, May 2019.

\bibitem{PhysRevA.60.1888}
Micha\l{} Horodecki, Pawe\l{} Horodecki, and Ryszard Horodecki.
\newblock General teleportation channel, singlet fraction, and
  quasidistillation.
\newblock {\em Phys. Rev. A}, 60:1888--1898, Sep 1999.

\bibitem{NIELSEN2002249}
Michael~A Nielsen.
\newblock A simple formula for the average gate fidelity of a quantum dynamical
  operation.
\newblock {\em Phys. Lett. A}, 303(4):249--252, 2002.

\bibitem{coderelease}
Evan Peters and Prasanth Shyamsundar.
\newblock Qubit assignment using time reversal.
\newblock \url{https://github.com/peterse/qubit-assignment-time-reversal},
  2022.

\bibitem{Wallman_2016}
Joel~J. Wallman and Joseph Emerson.
\newblock Noise tailoring for scalable quantum computation via randomized
  compiling.
\newblock {\em Phys. Rev. A}, 94(5), Nov 2016.

\bibitem{du2020learnability}
Yuxuan Du, Min-Hsiu Hsieh, Tongliang Liu, Shan You, and Dacheng Tao.
\newblock Learnability of quantum neural networks.
\newblock {\em PRX Quantum}, 2:040337, Nov 2021.

\bibitem{mi2021information}
Xiao Mi, Pedram Roushan, Chris Quintana, Salvatore Mandrà, Jeffrey Marshall,
  Charles Neill, Frank Arute, et~al.
\newblock Information scrambling in quantum circuits.
\newblock {\em Science}, 374(6574):1479--1483, 2021.

\bibitem{breuer2002}
H.P. Breuer and F.~Petruccione.
\newblock {\em The Theory Of Open Quantum Systems}.
\newblock Oxford University Press, 2002.

\bibitem{viola1999dynamical}
Lorenza Viola, Emanuel Knill, and Seth Lloyd.
\newblock Dynamical decoupling of open quantum systems.
\newblock {\em Phys. Rev. Lett.}, 82(12):2417, 1999.

\bibitem{PhysRev.80.580}
E.~L. Hahn.
\newblock Spin echoes.
\newblock {\em Phys. Rev.}, 80:580--594, Nov 1950.

\bibitem{2017_zne_a}
Kristan Temme, Sergey Bravyi, and Jay~M. Gambetta.
\newblock Error mitigation for short-depth quantum circuits.
\newblock {\em Phys. Rev. Lett.}, 119(18), Nov 2017.

\bibitem{2017_zne_b}
Ying Li and Simon~C. Benjamin.
\newblock Efficient variational quantum simulator incorporating active error
  minimization.
\newblock {\em Phys. Rev. X}, 7(2), Jun 2017.

\bibitem{botea2018complexity}
Adi Botea, Akihiro Kishimoto, and Radu Marinescu.
\newblock On the complexity of quantum circuit compilation.
\newblock In {\em Eleventh annual symposium on combinatorial search}, volume~9,
  pages 138--142, 2018.

\end{thebibliography}

\end{document}